\documentclass[useAMS,usenatbib,fleqn]{mnras}

\usepackage[T1]{fontenc}
\usepackage[utf8]{inputenc}
\usepackage{multicol}
\usepackage{times,amsmath,amssymb,longtable,breqn}
\usepackage[varg]{txfonts}
\usepackage{grffile} \usepackage{graphicx}
\usepackage{color}
\usepackage{xcolor}
\usepackage{ulem}
\usepackage{hyperref}
\usepackage{comment}

\newcommand\Msun{M_\odot}

\newcommand\rs[1]{_\mathrm{#1}}

\newcommand\Esn{E\rs{sn}}

\newcommand\Mej{M\rs{ej}}
\newcommand\rhoej{\rho\rs{ej}}
\newcommand\vej{v\rs{ej}}
\newcommand\Pej{P\rs{ej}}
\newcommand\Ppwn{P\rs{pwn}}

\newcommand\Rch{R\rs{ch}}
\newcommand\tch{t\rs{ch}}

\newcommand\tbegrev{t\rs{beg,rev}}

\newcommand\g{$\gamma$}

\newcommand\LL{\left(}
\newcommand\R{\right)}


\begin{document}
\label{firstpage}
\pagerange{\pageref{firstpage}--\pageref{lastpage}}

\title[Modeling the $\gamma-$ray PWNe population in our Galaxy]{Modeling the $\gamma-$ray Pulsar Wind Nebulae population in our Galaxy}
\author[M. Fiori et al.]
{M. Fiori$^{1,2}$,\thanks{michele.fiori@unipd.it}
B. Olmi$^{3,4}$,\thanks{barbara.olmi@inaf.it}
E. Amato$^{4,5}$, R. Bandiera$^{4}$, N. Bucciantini$^{4,5,6}$, L. Zampieri$^{2}$,\newauthor A. Burtovoi$^{4}$ \\
$^{1}$ Dipartimento di Fisica e Astronomia, Universit\`a  di Padova, Via F. Marzolo 8, I-35131, Padova, Italy \\
$^{2}$ INAF - Osservatorio Astronomico di Padova, Vicolo dell'Osservatorio 5, I-35122, Padova, Italy\\
$^{3}$ INAF - Osservatorio Astronomico ``Giuseppe S. Vaiana'', Piazza del Parlamento 1, I-90134 Palermo  Italy \\
$^{4}$ INAF - Osservatorio Astrofisico di Arcetri, Largo E. Fermi 5, I-50125 Firenze, Italy \\
$^{5}$ Dipartimento di Fisica e Astronomia, Universit\`a degli Studi di Firenze, Via G. Sansone 1, I-50019 Sesto F. no (Firenze), Italy \\
$^{6}$ INFN, Via G. Sansone 1, I-50019 Sesto F. no (Firenze), Italy \\}
\date{}
\maketitle

\pubyear{2021}

\begin{abstract}
Pulsar wind nebulae (PWNe) represent the largest class of sources that upcoming \g-ray surveys will detect.  
Therefore, accurate modelling of their global emission properties is one of the most urgent problems in high-energy astrophysics.
Correctly characterizing these dominant objects is a needed step to allow \g-ray surveys to detect fainter sources, investigate the signatures of cosmic-ray propagation and estimate the diffuse emission in the Galaxy.
In this paper we present an observationally motivated construction of the Galactic PWNe population.
We made use of a modified one-zone model to evolve for a long period of time the entire population. The model provides, for every source, at any age, a simplified description of the dynamical and spectral evolution. The long term effects of the reverberation phase on the spectral evolution are described, for the first time, based on physically motivated prescriptions for the evolution of the nebular radius supported by numerical studies. This effort tries to solve one of the most critical aspects of one-zone modeling, namely the typical overcompression of the nebula during the reverberation phase, resulting in a strong modification of its spectral properties at all frequencies.
We compare the emission properties of our synthetic Pulsar Wind Nebulae population with the most updated catalogues of TeV Galactic sources. %, namely the TeVcat and the H.~E.~S.~S. Galactic plane survey. 
We find that the firmly identified and candidate PWNe sum up to about 50\% of the expected objects in this class above threshold for detection. %Aside from the selection in flux, the detected objects are especially close, and hence compact in terms of angular size, which makes them more easy to identify. While not complete in terms of flux, the H.~E.~S.~S. sample is found, however, to be representative of the entire synthetic population in terms of the correlation between the Pulsar  TeV luminosity, $L_{1-10{\rm TeV}}$, and distance and between $L_{1-10{\rm TeV}}$ and pulsar spin-down power.
%When looking at the number of sources above some threshold flux, instead, our study suggests that some of the unidentified sources, with fluxes in the range 1-10\% of the Crab flux, might be halos, not accounted for in this work, rather than PWNe. 
Finally, we estimate that CTA will increase the number of TeV detected PWNe by a factor $\gtrsim3$.
\end{abstract}


%%%%%%%%%%%%%%%%%%%%%%%%%%%%%%%%%%%%%%%%%%%%%%%%%%
\section{Introduction}
\label{sec:intro}
A pulsar wind nebula (PWN) is a relativistic bubble generated by the interaction of the wind from a rotating neutron star, the pulsar (PSR), with the surrounding ambient medium. In young systems the latter is formed by the debris from the explosion of the progenitor star, the supernova ejecta.
As the pulsar spins down, losing rotational energy, the largest part of it is converted into a relativistically expanding, magnetized outflow, mainly - if not totally - composed by electron-positron pairs:  the pulsar wind.
In order to match the boundary conditions of the slowly expanding supernova ejecta, this wind is forced to slow down at a strong magneto-hydrodynamic termination shock (TS), where the plasma is heated and potentially strong magnetic dissipation takes place (for a review see e.g.  \citealt{Gaensler_Slane06a}). 
The same shock is also thought to be the place for particle acceleration, with evidence for leptons accelerated up to PeV energies in the Crab nebula \citep{Arons:2012, Amato:2014,Buhler_Blandford14a}.

A PWN shines in a broad range of energies, from radio to \g-rays. The lower energy emission - up to about 100-200 MeV (for young PWNe, with magnetic fields in the 100 $\mu$G range) is typically the result of synchrotron radiation by particles interacting with the nebular magnetic field. The energy dependent size of many PWNe reflects the fact that higher energy particles have shorter life-times against radiation losses and thus survive on shorter distances from their injection location at the shock.
The synchrotron emission is often highly polarized \citep{Novick:1972,Weisskopf1978,Velusamy85a} in a way that suggests that the magnetic field in the inner regions is mostly toroidal \citep{Kennel_Coroniti84b} and ordered to a high degree. Finally, in terms of morphology, the well known X-ray \textit{jet-torus} shape identified in the inner region of several PWNe \citep{Weisskopf:2000, Helfand:2001,Pavlov:2003,Gaensler:2001,Gaensler:2002,Lu_Wang+02a} is interpreted as due to the anisotropic energy injection from the pulsar wind. 

At frequencies larger than the synchrotron cut-off, the PWN emits radiation via inverse Compton scattering (ICS) of electrons and positrons on the local radiation fields: the background of synchrotron photons; the cosmic microwave background (CMB); infrared (IR) thermal photons from the local dust; photons coming from  background stars.
At TeV energies, the major contributors to the ICS emission are the relatively low energy leptons responsible for radio-IR synchrotron emission. 
Since these have longer radiation lifetimes than the X-ray emitting particles,   
a PWN will then remain bright in Very High Energy (VHE) \g-rays even when the X-ray emission has completely faded away.
The life-time of a \g-ray PWN, between 100 GeV and 300 TeV, can be estimated to be $\sim50-100$ kyr.
Considering an expected birth rate of pulsars in the Galaxy of $\simeq$1/100 yrs \citep{Faucher2006}, one can naively estimate around $\sim500-1000$ \g-ray emitting PWNe.
This means that, among the many classes of Galactic \g-ray emitting sources, PWNe are likely to be the most numerous. As a consequence, their 
209
 identification/discrimination will be one of the biggest challenges in the analysis of the data obtained by the next generation of \g-ray instruments, such as the Cherenkov Telescope Array (CTA). In fact, most of the newly detected \g-ray PWNe will not have any associated lower energy emission to guide their identification.
Already in the H.E.S.S. Galactic Plane Survey (HGPS hereafter), out of 24 extended sources, for which a PWN has been invoked, only 14 have been firmly identified with known objects, with the remaining 10 having no clear counterpart at other wavelengths \citep{HESScoll:2018-PWN}.

Typically, the two preferred ways to identify a TeV source as a PWN are: (i) the detection of a spatially coincident PWN at lower energies \citep{Kargaltsev:2013}; (ii) the co-location of a pulsar in the TeV emitting area. 
The latter strategy is only applicable to a fraction of the sources, since the pulsar can be seen directly only when its beamed radiation intercepts our line of sight.
The multi-wavelength association is then usually preferred for a firm identification.
For statistical reasons, related to the larger number of older objects, most of the \g-rays from PWNe will come from evolved systems, whose X-ray emission has long faded away, and whose radio emission is weak and extended, difficult to detect on top of the background. Devising an effective strategy to identify these systems is a problem of increasing urgency as the operational phase of CTA approaches. Some efforts to develop reliable models to establish more safely the possible associations are starting to be made \citep{Olmi:2020}.

In this work we present the first attempt to reproduce the population of \g-ray emitting PWNe based on state-of-the-art modeling of the PWN evolution and on current knowledge of the pulsar population and associated SNRs.
The two main novelties of our approach with respect to analogous efforts in the literature are: 1) a revised, physically informed, one-zone treatment of the evolution and radiation properties of these systems; 2) the adoption of a population of Galactic pulsars selected based on \g-ray data, so as to be representative of young enough objects to power observable nebulae. 
It is worth pointing out that we simulate the evolution of each system individually, in contrast with previous studies (e.g. \citet{HESScoll:2018-PWN}), where a single "average" source was considered and its parameters varied so as to match the observed population of TeV emitting PWNe.
We compare the predictions for the \g-ray emission from our entire synthetic population of PWNe with observational results from the HGPS and extract global trends. 
While also VERITAS \citep{Aliu:2013,Aliu2014,Mukherjee2016} and MAGIC \citep{Magic:2010, Magic2016} have greatly contributed to improve our knowledge of PWNe at TeV photon energies, our choice of the HGPS as a reference is due to the fact that this provides the most complete available survey of these sources, comprehensive of all but one (J1831-098) of the PWNe reported in TeVcat\footnote{\url{tevcat2.uchicago.edu}}.
Before concluding, we also  briefly discuss how the advent of CTA is expected to further improve our current knowledge of these systems. 

The paper is organized as follows: in Sec.~\ref{sec:pwne_initial} we describe and discuss the assumptions at the basis of the PWN population model; in Sec.~\ref{sec:pwne} we present the numerical tools used to generate the synthetic population and its complete spectral evolution; in Sec.~\ref{sec:results} we discuss  our results and compare with observations available from the HGPS.
Finally, in  Sec.~\ref{sec:conclusions} we draw our conclusions and comment on possible further developments of the model.

%%%%%%%%%%%%%%%%%%%%%%%%%%%%
\section{Generating the population}
\label{sec:pwne_initial}
%%%%%%%%%%%%%%%%%%%%%%%%%%%
An observationally-motivated model of the \g-ray emitting PWNe in the Galaxy must take into account the following aspects:
\begin{enumerate}
    \item the distribution of core collapse SNRs in the Galaxy;
    \item a population of pulsars able to account for the formation of newborn PWNe;
    \item the association of each pulsar with a core collapse SNR;
    \item the evolution of PWNe from  birth to the late stages.
\end{enumerate}

For the synthetic population of Galactic SNRs we have used the one presented in  \citet{Cristofari2017}, which had been optimized for reproducing the \g-ray SNRs of the Galaxy.
%---- SNRs---------
In that work, the authors located core collapse (CC hereafter) SNRs according to the spatial distribution of Galactic pulsars as modeled by \citet{Faucher2006} (FGK06 hereafter). The rate of supernova explosions is taken to be 3 per 100 years. Each remnant has an associated energy $\Esn=10^{51}$ erg, which is generally assumed as a representative value (see however \citealt{Hamuy:2002, Nadyozhin:2003, Zampieri2003, Muller:2017} for the variability of CC supernova energetics).
\citet{Cristofari2017} considered a wide range of densities for the ISM ($10^{-5}$-- 10 particles cm$^{-3}$), with the spatial distribution taken from \citet{Nakanishi:2003,Nakanishi:2006}.
%
%%%%%%%%%%%%%%%%%%%%%
\begin{figure*}
\centering
	\includegraphics[width=.545\textwidth]{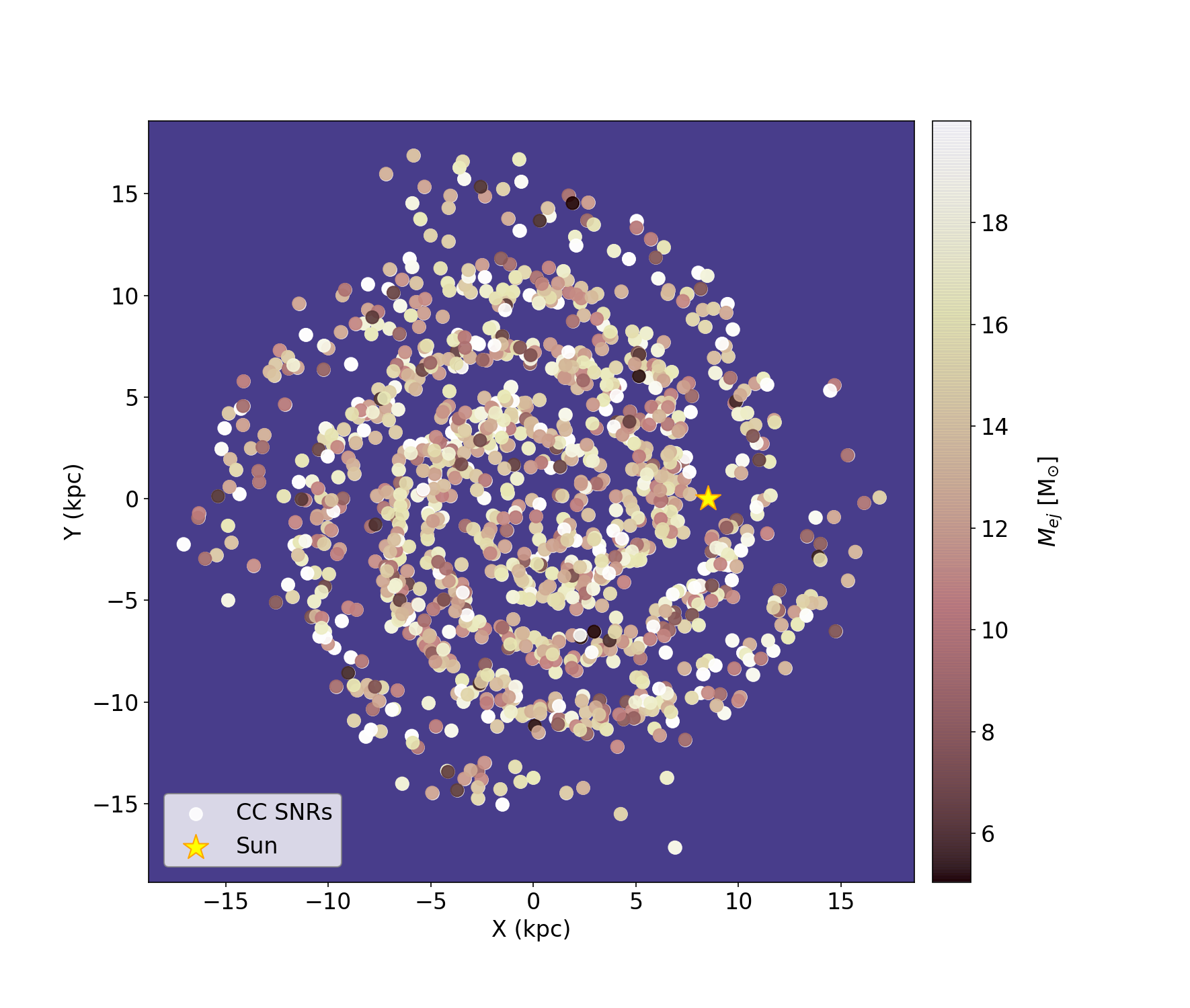}\,
	\hspace{-1.5cm}
	\includegraphics[width=.46\textwidth]{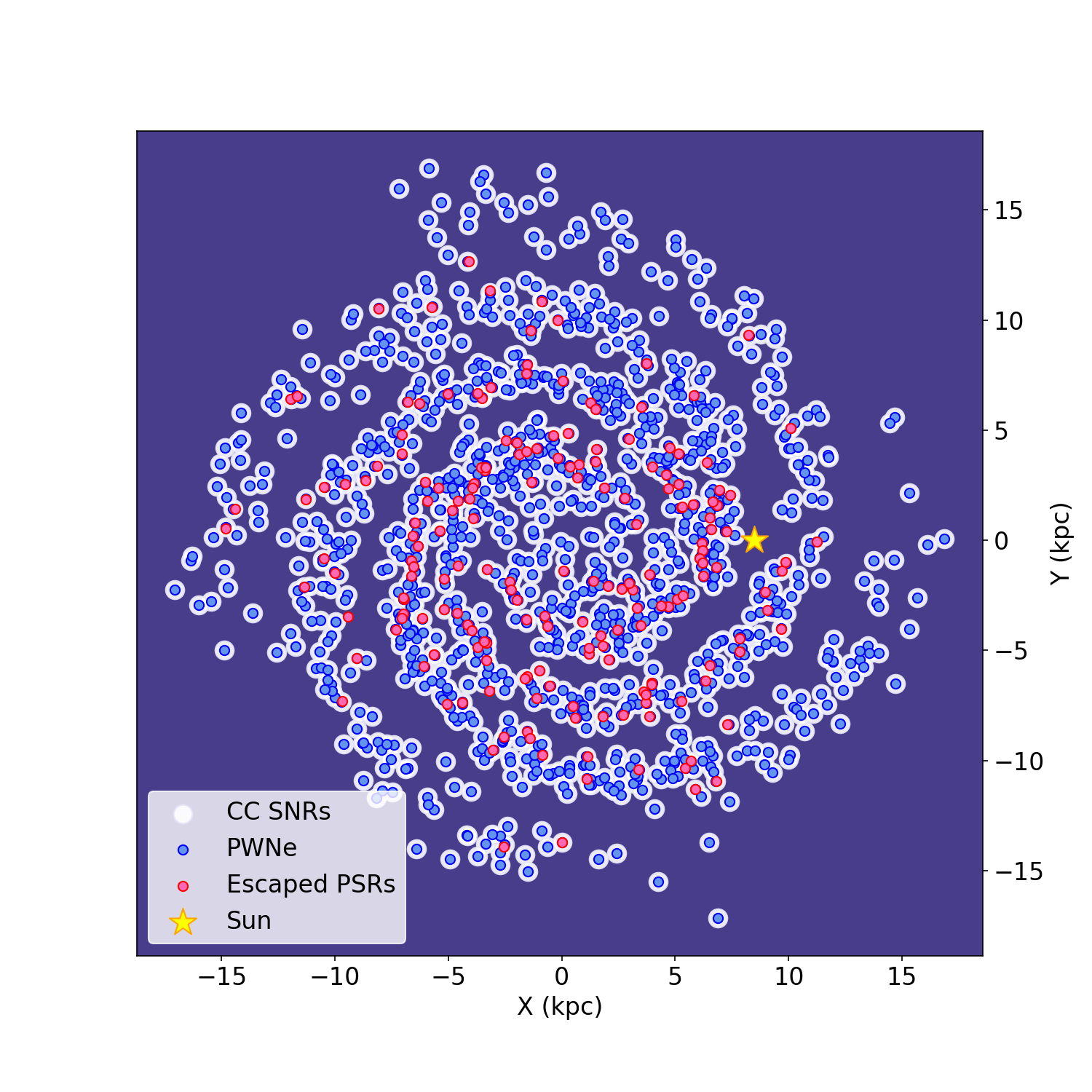}
    \caption{\textit{Left panel}: Reference population of the Galactic core collapse (CC) SNRs. The distribution of the ejecta masses (in units of solar mass) is color coded as indicated in the colorbar.
    \textit{Right panel}: The PWNe population on top of CC SNRs. Pink circles mark runaway PSRs that, at the end of their evolution, have left their parent SNR bubble, while blue circles are for systems that remain inside the remnant during the entire evolution. 
    In both plots the Sun position is marked with a yellow star and the Galactic coordinates are expressed in kpc.}
    \label{fig:SNR+PWNdist}
\end{figure*}
%%%%%%%%%%%%%%%%%%%%%
CC SNR masses are taken from a Gaussian distribution peaking at $13\,\Msun$, with $\sigma=3\,\Msun$. The distribution is cut at a minimum value of $5\,\Msun$ and the maximum one is imposed to be $20\,\Msun$, with those systems having $\Mej>20\Msun$ reset to $\Mej=20\,\Msun$ \citep{Smartt:2009}.
The latter choice is somewhat arbitrary, but given the lack of information on the distribution of $\Mej$ among the PWNe population, we decided to adopt the simplest approach to enforce the upper limit suggested by \citet{Smartt:2009}. In any case, this choice affected  a relatively small fraction of the entire sample ($\lesssim 8\%$).
The spatial distribution of the considered CC SNRs in the Galaxy, as well as ejecta mass distribution, is shown in the left-hand panel of Fig.~\ref{fig:SNR+PWNdist}. 
%

% --- pulsars---
To produce the synthetic population of Galactic PWNe we have then to associate a pulsar to each CC SNR of the sample.
A possibility would be to use the pulsar population described in FGK06, defined on the basis of the observations of Galactic radio emitting pulsars. 
This population, however, is clearly dominated by evolved objects (namely PSRs with characteristic ages typical of rotation powered radio pulsars , roughly in the range 1-20 Myr), by construction, and thus appears not to be the best choice in the present context, where the intent is that of simulating PWN powering, and hence young (age $\lesssim 1$ Myr), pulsars.

We considered more appropriate for the present scope to adopt the population of the \g-ray emitting pulsars that trace mostly young neutron stars. In particular, we chose the one described in \citet{Watters:2011} (WR11 hereafter), also similar to the one recently presented by \citet{Johnston:2020}.
To evaluate the impact of this choice on the final results, we run our model with different recipes for the pulsar population and compared the results with available \g-ray data.
We will discuss this point in more detail later in this section.

%%%%%%%%%%%%%%%%%%%%%
\begin{figure*}
\centering
\includegraphics[width=.45\textwidth]{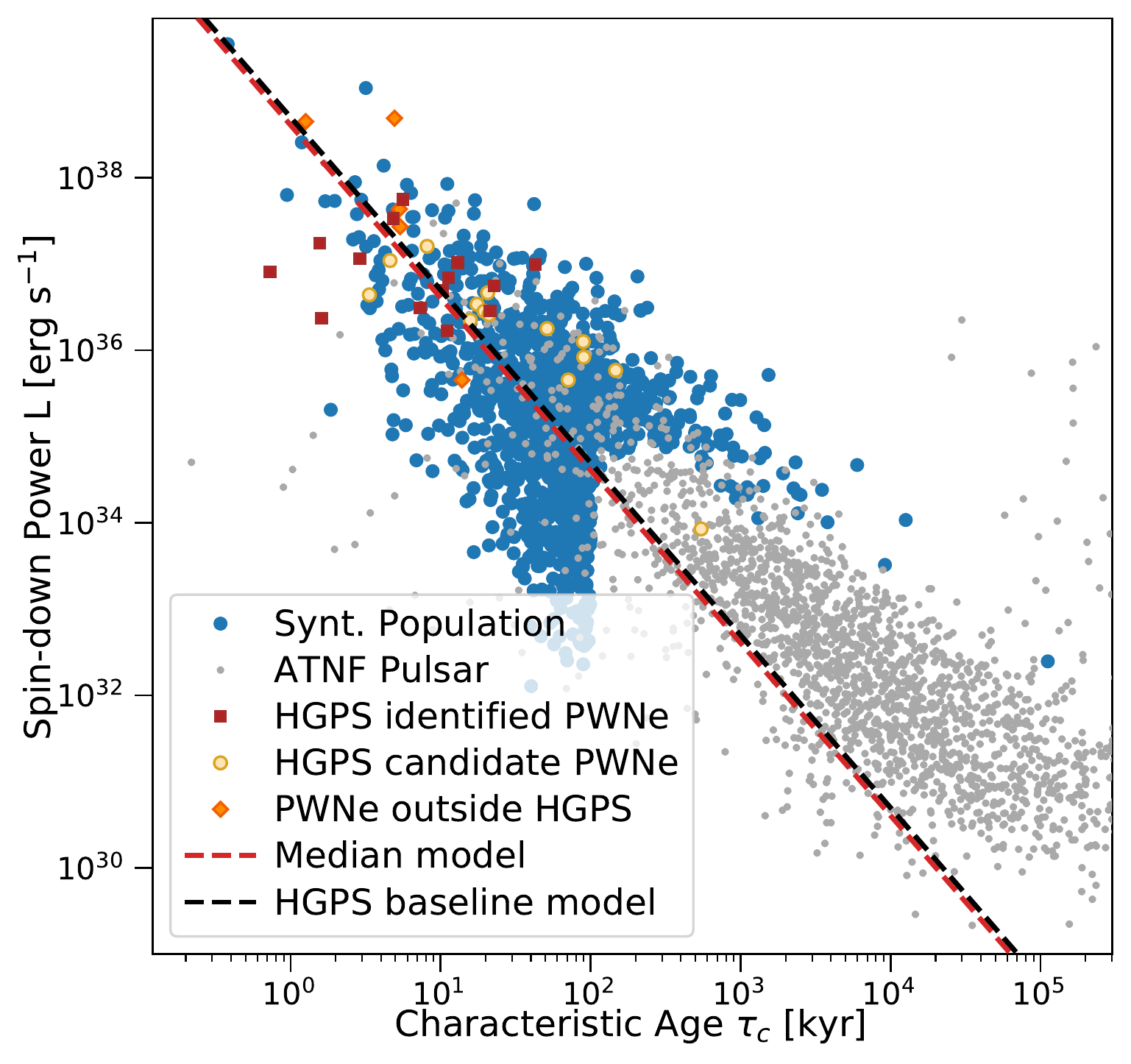}   	\includegraphics[width=.45\textwidth]{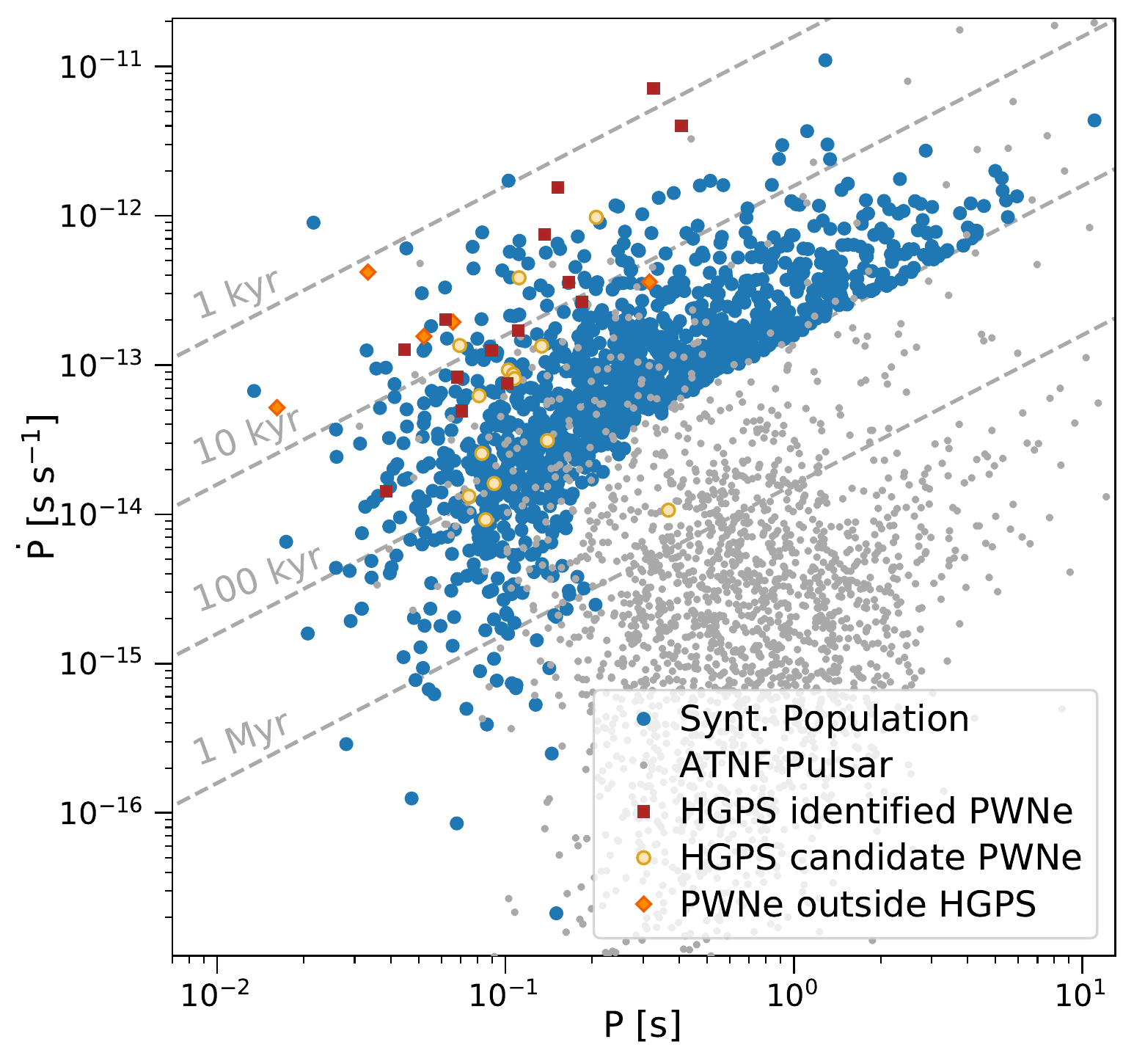}                 
\caption{Some characteristic quantities of the chosen population of pulsars (blue circles), to be directly compared with similar plots from \citet{HESScoll:2018-PWN} (see e.g. their Fig. 2). The other symbols in the plot represent: PSRs powering PWNe that are firmly identified in the HGPS (red squares); PSRs possibly associated with candidate HGPS PWNe (yellow circles); the 5 PSRs powering nebulae that are included in the HGPS plots as {\it outside HGPS PWNe} (orange diamonds), because the associated PWNe are outside the HGPS field, or have not been analysed with the HGPS pipeline (see \citet{HESScoll:2018-PWN} for details); all other pulsars from the ATNF catalogue (grey dots). 
\textit{Left Panel:} distribution of the pulsars in the characteristic age--luminosity diagram. Dashed lines represent the median of our distribution (red) and the HGPS baseline model (black). \textit{Right Panel:} distribution of synthetic pulsars in the spin-down period--spin-down period derivative ($P$-$\dot{P}$) diagram. Dashed lines indicate different characteristic ages of the sources in the population.}\label{fig:pPOP}
\end{figure*}
%%%%%%%%%%%%%%%%%%%%%

Except for the distribution of the initial pulsar spin  periods ($P_0$), the populations in FGK06 and WR11 are very similar.
The magnetic field is modeled with a log-normal distribution centered at $\log_{10}(B/\rm{G})=12.65$ and a spread of $\sigma_{\log_{10}{B}}=0.55$.
In both FGK06 and WR11 the pulsar braking index is assumed to be the one relative to a rotator with pure dipole spin-down, namely  $n=3$. 
Actually, this parameter shows significant variations \citet{Parthasarathy:2020}, not fully understood.
For the present work we assumed $n=3$, as commonly done in the literature and for lack of a better prescription, but we are aware that deviations from this value might introduce quite significant differences in the final pulsar population properties.
In FGK06 the initial spin-down period is centered at the mean value $\langle P_0 \rangle=300$ ms, with a spread $\sigma_{P_0}=150$~ms, while in WR11 it is centered at 50 ms, with a spread equal to the mean, and truncation at 10~ms. 
The associated probability density function \textit{(PDF)}, for $P_0>10$~ms is then:
\begin{equation}
\label{eq:P0}
\mathcal{P}(P_0)\propto e^{-(P_0- \langle P_0\rangle)^2/( 2 \langle P_0\rangle^2)}\,,
\end{equation}
with $P_0$ the initial spin-down period.
It is worth mentioning that, contrary to the radio population that has spun down enough to retain little memory of the young phase, the \g-ray population is instead sensitive to variations of the initial spin-down period, and different choices may lead to quite different final populations \citep{Watters:2011, Johnston:2020}. In the context of the present work, the initial pulsar period is particularly relevant because that is what sets the total available spin-down energy, $E_{\rm rot}=2 \pi^2 I _{\rm PSR}/P_0^2$ (with $I_{\rm PSR}$ the pulsar moment of inertia), and hence determines the evolution of the system at late times.

The validity of our assumption that \g-ray pulsars well represent the young population of pulsars powering PWNe in the Galaxy has been verified by running multiple simulations of the entire population, varying the distribution of initial spin-down periods $P_0$. 
In particular, we considered four different possibilities: pure FGK06 (centered at 300 ms), pure WR11 (centered at 50 ms) and two intermediate populations, the first one with $P_0$ centered at 80 ms and the second with $P_0$ centered at 120 ms. For all these cases, we computed the \g-ray emission and then compared the results with the available \g-ray data, as taken from the gamma-cat catalogue.\footnote{\url{gamma-cat.readthedocs.io}}
We found that the PWN population based on the WR11 pulsar distribution is the one that best reproduces the available observations. In all the other cases the resulting \g-ray flux exceeds the observed one. 

We associate a three-dimensional kick velocity to each pulsar of the population, considering the double-sided exponential velocity distribution model of FGK06, with a mean value of $\sim 380$ km s$^{-1}$. Since SNRs are in decelerated expansion in the ambient medium, a large fraction of all the pulsars are actually expected to escape from the bubble of their parent SNR on time scales comparable with the final age considered for this work, and then to form bow shock pulsar wind nebulae (BSPWNe) shaped by the interaction with the ISM \citep{Bucciantini2002,van-der-Swaluw:2003,Bucciantini:2005,Vigelius:2007,Kargaltsev:2017,Barkov:2019,Olmi:2019,Toropina:2019}.
BSPWNe are perfect locations where to look for efficient particles escape \citep{Olmi:2019b}, and associated TeV halos \citep{Abeysekara:2017, Sudoh:2019}. %
The population of PSRs and associated SNRs is generated using a Monte-Carlo technique and assuming no correlation among the various parameters describing the system. This produces the initial synthetic population of PWNe shown in the left panel of Fig.~\ref{fig:SNR+PWNdist}, where we highlight also the PSRs escaped from their parent remnants at the end of the simulation. 

Considering a pure dipole braking, the spin-down luminosity of the pulsar, corresponding to the energy flux into the PWN, at the generic age $t$ is:
\begin{equation}\label{eq:edot}
L(t) = \frac{L_0}{ \LL 1 + t/\tau_0  \R^{2}}=4\upi^2 I \LL \frac{\dot{P}}{P^3}\R \,,
\end{equation}
with $L_0$ the initial spin-down luminosity, $\tau_0$ the spin-down time, $I$ the pulsar momentum of inertia (usually assumed to be $I=10^{45} $ g cm$^2$) and $\dot{P}$ the time derivative of the spin period $P(t)$.
The spin-down time at birth is given by:
\begin{equation}\label{eq:t0}
	\tau_0 = \frac{4\upi^2 R_*^6 B_0^2}{3 I c^3 \dot{P_0}^2}\,,
\end{equation}
where $R_*$ is the neutron star radius and $B_0$ its magnetic field at the pole. 
The characteristic age of the pulsar is:
\begin{equation}\label{eq:tauC}
    \tau_c = \frac{P}{2\dot{P}}= \tau_0 + t_{\rm{age}}\,,
\end{equation}
and it provides a good approximation of the real age $t_{\rm{age}}$ only for $\tau_0 \ll t\rs{age}$.

In Fig.~\ref{fig:pPOP} we compare our synthetic population both with all the pulsars from the ATNF catalogue\footnote{\url{www.atnf.csiro.au/people/pulsar/psrcat/}}, and with those having an associated PWN, either from the HGPS or from other independent observations \citep{outHGPS_crab_1, outHGPS_crab_2, outHGPS_g0901, outHGPS_3c58, outHGPS_n175_1, outHGPS_n175_2, outHGPS_cta1}.
Our population is limited to systems with age younger than $t_{\rm{end}}=10^5$ yr, which translates into a cut in our synthetic sample, as can be clearly seen in Fig.~\ref{fig:pPOP}. We have verified however, by varying the final age, that this does not affect the results of the present study, given that for older PSRs the $\gamma$-ray luminosity of the associated PWN is negligible.
In the distribution of luminosities versus characteristic ages, we also compare the baseline model from the PWNe analysis in the HGPS (red dashed line - see Eq. 5 of \citealt{HESScoll:2018-PWN}) with that computed using the median values of $L_0$ and $\tau_0$ from our synthetic population.  It is noteworthy that the two appear very close together, despite the simplified assumption, and limited number of objects, used to build the baseline model in \citet{HESScoll:2018-PWN}.
%
%%%%%%%%%%%%%%%%%%%%%%%%%%%%%%%%%%%%%%%%%%%%%%%%%%
\section{Evolving the population}
\label{sec:pwne} 
Typically, our population contains $\sim 1300$ sources, which are all evolved up to the final age of $t_{\rm{end}}=10^5$ yr. To model the spectral and dynamical evolution of each PWN, we adopt the so-called one-zone approach \citep{Venter2007, Zhang2008, qiao2009, Gelfand:2009, fang&zang2010, Tanaka2010, tanakatakahara2011,Bucciantini_Arons+11a, Martin:2012, Martin2016, Torres2014, Zhu2018, vanRensburg2018, Fiori:2020}.  
This approach has been widely and successfully applied to systems of different ages, and allows one to rapidly model the full evolution, even for very old systems. 
With current computational capabilities, this is not feasible
with more sophisticated 1D or multi-D approaches based on hydrodynamic (HD) or even magneto-hydrodynamic (MHD) numerical simulations \citep{Blondin:2001,van-der-Swaluw:2001,Komissarov_Lyubarsky04a, Del-Zanna_Volpi+06a, Porth:2014, Olmi:2016, Olmi:2020}.
In addition, there are physical reasons to believe that one-zone models might provide a better description of the system than reduced dimensionality MHD simulations: in fact, 1D and 2D MHD models fail to capture the important role that turbulence and mixing are bound to play in the late PWN dynamics (see e.g. the discussion on the differences of 2D and 3D dynamics in \citealt{Porth:2014} or \citealt{Olmi:2016}).

In one-zone models the PWN is treated as a homogeneous system, whose evolution is governed by the interaction with the surrounding SNR and by particles and energy losses (both adiabatic and radiative).  
In particular, the PWN radius is treated as coincident with that of the thin and  massive shell of swept-up material that accumulates at the PWN boundary as it expands, initially in the un-shocked ejecta and later on in the SNR shell \citep{van-der-Swaluw:2001,Bucciantini_Blondin+03a,Gelfand:2009,Martin:2012}.
The one-zone approach was used multiple times in the past, and it was proved to give robust results, at least in describing the first phase of the evolution, when the PWN expands with mild-acceleration within the freely expanding SNR.
Recently, \citet{Bandiera:2020} have shown that those models must be used with caution when addressing longer evolution time-scales and passing through the phase known as \textit{reverberation}.
The reverberation phase begins when the SNR reverse shock (RS hereafter), that moves from the border of the SNR towards the center, reaches the boundary of the PWN bubble. 
Depending on the energetics of the PWN and the pressure in the SNR, this interaction may cause a series of contractions and re-expansions of the PWN, which possibly affect its spectral and morphological properties.
The effects of the modifications induced by  the reverberation are usually quantified through the so called \textit{compression factor} ($c_f$), namely the ratio between the maximum radius of the PWN -- reached in the early reverberation phase, before the PWN starts contracting -- and the minimum one -- namely that when the PWN is maximally compressed.
A dramatic modification of the spectral properties of a PWN, called super-efficiency, was described by \citet{Torres2019}, who found very large compression factors, up to $c_f>1000$. In these extreme cases, the particle heating and the enhancement of the nebular magnetic field are so huge as to cause catastrophic synchrotron losses, strongly reducing the number of particles available for ICS emission at late times.
More recently, \citet{Bandiera:2020} have shown that this super-efficiency phase might actually be only expected for a small fraction of the PWNe population, while most of the systems will experience a maximum compression of a few tens, that will not reflect in such a drastic modification of the spectral properties at all wavelengths.
The possibility of strong compressions and later re-expansions during the reverberation phase had been first criticized by \citet{Bucciantini_Arons+11a} and is the subject of an ongoing study of some of the authors of the present manuscript.
Current results indicate that the thin-shell approximation is not accurate enough when looking at the PWN properties close to the reverberation phase, and leads to overestimate the compression (\citealt{Bandiera:2020} and Bandiera et al. 2021, in preparation).

Of course, a realistic and detailed description of the PWN-SNR co-evolution  requires much more sophisticated modeling. 3D MHD simulations offer the minimum degree of complexity that still allows to account for essential phenomena, such as magnetic instabilities and dissipation (which may reduce the magnetic field enhancement during compression) and fluid instabilities (like Rayleigh-Taylor's) that might develop at the contact discontinuity and partly or completely disrupt the thin shell \citep{Blondin:2001, Bucciantini:2004, Porth:2014a,Kolb:2017,Olmi:2020}, hence strongly reducing the compression during the reverberation phase.
However a 3D MHD investigation, up to very late times, is beyond current computational possibilities.
3D models of young PWNe have shown to require millions of CPU hours and months of continuously running computations for reproducing only a limited part of their evolution \citep{Porth:2014, Olmi:2016}.
They have been then only used to investigate a limited number of sources.
Moreover these  models still lack of a consistent treatment of radiative losses, generally traced in the post-processing and not directly linked to the evolution (see e.g. \citealt{Komissarov:2004,Del-Zanna:2006,Olmi:2013}).

On the other hand, MHD simulations of reduced dimensionality are likely to provide an inaccurate description of the internal dynamics of the system (see e.g. the discussion in \citealt{Olmi:2020}), which in this context reflects in a poor description of the evolution during the most dramatic phases of the interaction between the PWN and the SNR reverse shock.
In light of recent and current investigations (\citealt{Bandiera:2020, Bandiera:2021a} and Bandiera et al. in preparation), we believe that one-zone models, with complete neglect of the internal dynamics and an informed prescription for the evolution of the nebular radius, provide more reliable results for the long term evolution of the system than reduced dimensionality MHD models.

\subsection{Dynamical evolution}
\label{sec:pwne_dyn}
The evolution of a PWN can be roughly divided into three distinct stages: (i) the free expansion phase, (ii) the reverberation phase, (iii) the relic stage.
In this work, we computed the evolution using the numerical code described in \citet{Fiori:2020}, with some modifications listed below to adapt it to the present problem.
Characterization of the different systems is made simpler by the introduction of some adimensional quantities. 
Let us consider the characteristic radius ($R_{\rm ch}$), time ($t_{\rm ch}$), and luminosity ($L_{\rm ch}$) of a SNR as defined by \citet{Truelove1999}:
\begin{eqnarray}
\!\!\!\!\!\!\!\!\!\!R_{\rm{ch}}\!\!\! &=&\!\!\! \Mej^{1/3}\rho_0^{-1/3}\,,	\\
\!\!\!\!\!\!\!\!\!\!t_{\rm{ch}}\!\!\!&=&\!\!\!
\Esn^{-1/2}\Mej^{5/6}\rho_0^{-1/3}\,,	\\
\!\!\!\!\!\!\!\!\!\!L_{\rm{ch}}\!\!\!&=&\!\!\!
\Esn/t_{\rm{ch}}\,.
\label{eq:TMC}
\end{eqnarray}
Here $\Mej$ is the mass of the supernova ejecta, $\rho_0$ the mass density of the medium in which the SN expands, $\Esn$ the energy of the supernova explosion.
With these scalings, one can define dimensionless quantities for the characteristic time and luminosity of the pulsar: 
\begin{eqnarray}
\tau^*\equiv \tau_0/t_{\rm{ch}} \label{eq:chart}\\ 
L^*\equiv L/L_{\rm{ch}}\,,
\label{eq:char}
\end{eqnarray}
and the energy released by the pulsar can be then parametrized by the product: $(L^* \tau^*)$.

%\item % --- Free expansion phase ----
\subsubsection{Free expansion phase}

In the first phase, the PWN expands with a mild acceleration in the freely expanding ejecta of the parent SNR. 
As mentioned before, this phase has been largely investigated with different approaches: from one-zone models to multi-dimensional HD and MHD simulations, with the latter often devoted to specific objects \citep{Reynolds:1984,van-der-Swaluw:2001, Bucciantini_Blondin+03a,Komissarov:2004,Del-Zanna:2006,Porth:2014,Olmi:2016}.
In the one-zone approach the thin shell of swept-up material at the PWN boundary $R$, evolves following mass conservation.
The evolution of the nebula is then described with the conservation of the shell mass $M$:
\begin{equation}\label{eq:thin-shell1}
\frac{dM(t)}{dt} = 4 \upi R^2(t) \rhoej(R,t)  [v(t)-\vej(R,t)]\,,
\end{equation}
and the shell momentum $M v(t)$:
\begin{equation}\label{eq:thin-shell2}
\frac{d}{dt}[M(t)v(t)]  =  4 \upi R^2(t) \left[  \Ppwn(t) - \Pej(R,t) \right] + \frac{dM(t)}{dt} \vej(R,t) \,,%F(t)\,, 
\end{equation}
where $v(t)=dR(t)/dt$, while $\rhoej$ and $\vej$ are the mass density and velocity of the homologously expanding SNR ejecta (for which we assume a core-envelope profile as in \citealt{Gelfand:2009}). In the momentum equation the force acting on the shell is given by the pressure difference between the PWN ($\Ppwn$) and the ejecta ($\Pej$), plus the contribution to the variation of momentum of the material swept-up from the ejecta.
Analytic approximations of the solutions, which have also been used to set the initial conditions for the reverberation phase, are given in Appendix~\ref{app:A}.

The pressure in the relativistic and homogeneous bubble that approximates the PWN is given by the sum of the magnetic pressure and the pressure of the relativistic particles, whose energetic evolves taking into account energy injection from the PSR as well as adiabatic and radiation losses (see Sect.~\ref{sec:pwne_sp})

\subsubsection{Reverberation phase}
\label{sec:reverberation}
The free expansion phase ends when the RS reaches the PWN boundary.
From that moment on, the PWN and the SNR shell start to interact directly and, as discussed previously, this generally causes a contraction of the PWN itself. 
The modelling of this phase is rather complex. Different {\it ad hoc} assumptions and prescriptions have been adopted, in the literature,  to describe the evolution of the swept-up mass or of the pressure within the SNR shell \citep[see e.g.][]{Bucciantini_Arons+11a}. None of these, however, was ever tested against a complete and representative set of numerical results. Here we improve on this limitation, by using a semi-analytic prescription for $P_{\rm SNR}(t)$ derived from a fit to the results of extensive 1D HD simulations performed with PLUTO \citep{Mignone:2007}. We considered the 
hydrodynamic evolution of the PWN-SNR interaction (in the absence of radiation losses) for a set of about 30 PWN-SNR systems \citep{poster}, that are supposed to cover the parameter space of our investigation as widely as possible. 
Based on the output of these simulations, we found that the one-zone model could well reproduce the compression when adopting the following time-dependence for the pressure of the SNR shell:
%%%
\begin{eqnarray}
P\rs{SNR}(t)\,\frac{\Rch^{3}}{\Esn}\!\!\!\!&=&\!\!\!\!0.00140+\nonumber\\
&&\!\!\!\!\!\!\!\!\!\!\!\!\!\!\!\!\!\!\!\!\!\!\!\!
\left[0.0412+0.0214\lg(L^*\tau^*)+0.0030\lg(L^*\tau^*)^2\right] \times \nonumber\\
&&\!\!\!\!\!\!\!\!\!\!\!\!\!\!\!\!\!\!\!\!\!\!\!\!
\exp\left[-\delta t\left(4.21+3.04\lg(L^*\tau^*)+1.04\lg(L^*\tau^*)^2\right)\right]+\nonumber\\
&&\!\!\!\!\!\!\!\!\!\!\!\!\!\!\!\!\!\!\!\!\!\!\!\!
\left[0.7892 + 0.4802\lg(L^*\tau^*)+0.0754\lg(L^*\tau^*)^2\right]\times \nonumber\\
&&\!\!\!\!\!\!\!\!\!\!\!\!\!\!\!\!\!\!\!\!\!\!\!\!
\exp\left[-\delta t\left(121.94+77.96\lg(L^*\tau^*)+15.64\lg(L^*\tau^*)^2\right)\right],\qquad
\end{eqnarray}
%%%
where $\delta t \equiv (t-\tbegrev)/\tch$, and $\tbegrev$ is the time when the reverberation phase begins, while $\lg$ is the logarithm in base 10.
Assuming that this prescription provides a good approximation even in the presence of relevant radiation losses, during the reverberation phase we can use a revised version of Eq.~\ref{eq:thin-shell2}:
%%%
\begin{equation}\label{eq:thin-rev2}
M\rs{shell}\frac{d}{dt}[v(t)]  =  4 \upi R^2(t) \left[ \Ppwn(t) -P\rs{SNR}(t) \right] \,,%F(t)\,, 
\end{equation}
where now the shell mass is kept fixed to the value $M\rs{shell}$ reached at $t=\tbegrev$. For $L^*\tau^*\gg 1$, approximate pressure equilibrium between the PWN and the SNR holds without effective compression of the PWN.  
For $L^*\tau^*\ll 1$, the evolution consists of an inward motion of the shell driven by the SNR pressure, which is only contrasted by the shell inertia. The PWN pressure only manifests for a brief time once the system is strongly compressed.

This approach represents a substantial improvement with respect to simply assuming a pressure proportional to the Sedov solution \citep[see e.g.][]{Gelfand:2009,Torres2014} and we have found it adequate to the aims of the present work, namely to obtain predictions of the statistics of the broad-band emission, with a special focus on the TeV range. A more sophisticated analysis, with a larger set of 1D models, is underway and results will be published in a forthcoming paper.
\subsubsection{Relic phase: BSPWNe, old PWNe and leftover bubbles}
As discussed before, a fraction  of PSRs is bound to emerge from the progenitor SNR before our fiducial final time $t_{\rm end}=10^5$yr, due to the high average kick velocity that characterizes the pulsar population. The typical escape time can be estimated by matching the PSR displacement due to its kick velocity ($V_{\rm psr}$) with the size of the SNR in the Sedov-Taylor phase:

\begin{equation}
    t_{\rm esc} \simeq 725\; \mbox{kyr} \;\left[\left(\frac{E_{\rm sn}}{10^{51} \,\mbox{erg}}\right)
    \left(\frac{\rho_0}{1\, \mbox{part/cm$^{3}$}}\right)^{-1} \left(\frac{V_{\rm psr}}{100\, \mbox{km/s}}\right)^{-5} \right]^{\,1/3}\!\!\! .
\end{equation}
Considering the mean (median)
value of the PSR velocity distribution of 380 (330) km s$^{-1}$ and of the ISM number density of 0.7 (0.25) particles cm$^{-3}$, we obtain a mean (median) escape time of $t_{\rm esc}\simeq88$ (160) kyr.
Even taking into account that transition of SNRs to the radiative phase is expected at 35 (60)  kyr, the escape time only slightly reduces to $t_{\rm esc}\simeq 77$ (120) kyr.
Since $t_{\rm end}=100$ kyr, only a fraction of the sources will then escape the SNR by the end of the simulation.
For those systems with $t_{\rm esc}<t_{\rm end},$ the runaway PSR will give rise to the formation of a bow shock nebula. 
These nebulae, whose first examples were detected in H$_\alpha$ \citep{Chevalier1980,Kulkarni_Hester88a,Cordes:93,Bell_Bailes+95a,van-Kerkwijk_Kulkarni01a,Jones_Stappers+02a,Brownsberger:2014,Dolch:2016,Romani_Slane+17a}, more recently have been discovered and observed in X-rays and sometimes in radio \citep{Gaensler:2002,Gaensler2004,Arzoumanian_Cordes+04a, Chatterjee:2005,Yusef-Zadeh:2005,Hui_Becker07a,Kargaltsev2008,Hui:2008,Misanovic_Pavlov+08a,de-Rosa:2009,Ng:2010, De-Luca_Marelli+11a,Ng:2012,Marelli_De-Luca+13a,Jakobsen_Tomsick+14a,Auchettl:2015,Klingler_Rangelov+16a,Posselt_Pavlov+17a,Kargaltsev:2017,Kim:2020}. 
They are characterized by a cometary shape, with a tiny head typically of the order of $10^{16}$ cm, whose size is set by ram pressure balance between the PSR wind and the incoming (in the PSR frame) ISM, followed by a long tail opposite to the PSR motion, which can extend for very long distances up to a few pc.
Given their limited spatial extension and low residual luminosity \citep{Kargaltsev:2017}, bow shock nebulae will not probably be statistically relevant in \g-rays, and so far have not been detected \citep{HESScoll:2018-BSPWNe}. Runaway PSRs however, have recently been associated to extended TeV halos \citep{Abeysekara:2017,Sudoh:2019}, most likely due to escaping pairs \citep{Bykov:2017,Evoli:2018,Olmi:2019b,Di-Mauro:2020,Evoli:2021}. 
However, the formation and properties of \g-ray haloes are still poorly understood, and different interpretations lead to very different expectations in terms of the possible detection of these sources in the next future \citep{Sudoh:2019,Giacinti:2020}. The modelling of these complex sources is outside the scopes of the present work, so we simply keep trace of the position of escaped PSRs for possible future implementations. 
The fraction of PWNe escaped from their parent SNR at the end of the simulation is represented in the right panel of Fig.~\ref{fig:SNR+PWNdist}, where evolved PWNe are shown on top of the initial distribution of PWNe+SNRs. %

More relevant for the present work is the role of the relic bubbles of electrons injected in the nebula during the PSR history. Among these particles, the lowest energy ones, producing radio synchrotron radiation, will not have cooled down by the time the PSR escapes the SNR or moves outside of the original wind bubble (which usually happens eariler than $t_{\rm esc}$), 
so that these systems are possible sources of ICS \g-ray emission. 
The escape of the PSR from its original wind bubble, proven by the existence of highly asymmetric systems, is a problem with which all one-zone models struggle to deal \citep{Gelfand:2009}. Starting from the time the PSR escapes, the nebula of leftover electrons is treated as a relic subject to adiabatic expansion alone. We will come back to this point at the end of Section \ref{sec:results}, with a more quantitative discussion.
%
%%%%%%%%%%%%%%%%%%%%%%%%%%%%%%%%%%%%%
\subsection{Spectral evolution}
\label{sec:pwne_sp}
The spectral evolution of the PWN is calculated using the one-zone model implementation of \cite{Fiori:2020}. The evolution in time of the spectral energy distribution of the particles, $N(E,t)$, is given by:
\begin{equation}\label{p_transp}
	\frac{\partial N(E,t)}{\partial t} = Q(E,t) - \frac{\partial}{\partial E}\left[b(E,t)N(E,t) \right]- \frac{N(E,t)}{\tau_{\rm{esc}}(E,t)}\,,
\end{equation}
where $E$ is the particle energy, $Q(E,t)$ the source term and $\tau_{\rm{esc}}$ the characteristic time for particle to escape from the system. Finally, $b(E,t)$ is the particles energy loss rate and it takes into account all the different loss processes that particles may experience: synchrotron, ICS, and adiabatic. In addition, it depends on the magnetic field of the system, its evolutionary history and its location within the Galaxy.
The injection spectrum of the source term is assumed to be well described by a broken power law, as suggested by broadband observations of many PWNe \citep[see e.g.][]{Gaensler_Slane06a,Reynolds2017}:
\begin{equation}\label{eq:source}
	Q(E,t) = Q_0(t) \begin{cases} (E/E_b)^{-\alpha_1} & \mbox{if } E\leq E_b \\
	(E/E_b)^{-\alpha_2} & \mbox{if } E > E_b
\end{cases}\,,
\end{equation}
where $E_b$ is the break energy. Its distribution for the population of PWNe is modeled as log-normal, with a mean value of 0.28 TeV and a spread of 0.12 TeV (values based on the results by \citet{Bucciantini_Arons+11a, Torres2014}).
The time evolution of the normalization $Q_0(t)$ is computed assuming that the power of the injected particles is equal to a fraction of the PSR spin-down power $L(t)$:
\begin{equation}
(1-\eta)~L(t)=\int_{E_{\rm{min}}}^{E_{\rm{max}}} E^{\prime} Q(E^{\prime},t) dE^{\prime},
\end{equation}
where $\eta$ is the fraction of luminosity injected in the PWN in the form of magnetic field.
The minimum energy for the injected particles is not relevant for the \g-ray emission provided that $E_{\rm{min}}<E_b$; we then set it to $E_{\rm{min}}=10^3 m_e c^2$, with $m_e$ the electron mass. 
The maximum energy must be such that $E_{\rm{max}}>E_b$, and it is randomly varied imposing that it always stays well beyond the maximum energy achievable via acceleration, that associated with the PSR maximum potential drop $e[\dot{E}(t)/c]^{1/2}$, with $e$ the electron charge, as it is generally assumed in one-zone models \citep{Gelfand:2009, Torres2014}.
In particular, the maximum energy of emitting particles can be radiation limited, especially at early times, so that we always set $E_{\rm{max}}$ as the minimum between the theoretical limit and the radiation limit, computed as the balance between acceleration and synchrotron losses (following the approach in \citealt{de-Jager:1996}).
Finally, the power law indices $\alpha_1$ and $\alpha_2$ are varied in the range $1.0<\alpha_1<1.7$ and $2.0<\alpha_2<2.7$ \citep{Gaensler_Slane06a,Reynolds2017}.

As far as losses are concerned, we include the possibility that particles can leave the nebula as a result of diffusion, a possibility that is usually neglected in HD and MHD models of the nebular dynamics (see e.g. \citealt{Kennel:1984a}), where particle transport is assumed to be governed by advection at all energies). We describe particle escape as due to diffusion in the Bohm regime, the simplest scenario in the highly turbulent field that one expects in evolved systems:
$\tau_{\rm{esc}}(E,t)= 3 eB(t) R^2(t)/(E c)$ . ICS losses are computed using the Klein-Nishina cross-section \citep{Blumenthal:1970} and considering the interaction of the leptons in the PWN with different photon fields: synchrotron emitted photons, photons of the CMB, and those of near- and far- IR components.
The IR background field is modeled with a normal distribution centered on the value obtained from the GALPROP model for the ISM \citep{Porter:2017} at the position of each source. 
This choice follows results of previous works (see e.g. the discussion in \citealt{Torres2014}), showing that the IR background must be in many cases modified with respect to the GALPROP model expectation in order to correctly reproduce the PWN properties. Our approach is meant to account for these local modifications.
The magnetic field energy $W_B(t)$ is evolved following \citet{Gelfand:2009} and  \citet{Martin2016}, subject to the adiabatic expansion/contraction of the nebula and to the energy injection from the PSR:
\begin{equation}\label{eq:Bfield1}
	\frac{dW_B(t)}{d t}= \eta \dot{E}(t) - \frac{W_B(t)}{R(t)}\frac{d R(t)}{dt}\,,
\end{equation}
with $W_B(t)= \upi B^2(t)/(6\upi )  R^3$. The above equation integrates to:
\begin{equation}\label{eq:Bfield2}
	B(t)=\frac{\sqrt{6\eta}}{R^2(t)} \left[ \int_0^t \dot{E}(t^\prime)R(t^\prime) dt^\prime  \right]^{1/2}\!\!\!.
\end{equation}
For the systems that become relic, namely those for which the pulsar escapes the nebula at a time $t<t_{\rm{end}}$, we have introduced a threshold value for the magnetic field: $B_{\rm{sim}}(t)=max[B(t),B_{\rm{floor}}]$, where 
$B_{\rm{floor}}\equiv 5\;\mu$G. 
This value is slightly higher than the one expected in the ISM ($B_{\rm ISM}\approx 3\ \mu$G) and comparable to the magnetic field strength typically inferred from modeling of BSPWNe \citep{Olmi:2019}. The exact value of $B_{\rm floor}$ only impacts the ratio between ICS and synchrotron emission for relic nebulae. 
%%%%%%%%%
% Tabella random source
\begin{table}
%\scriptsize
\centering
\caption{Parameters of the source illustrated in Fig.~\ref{fig:spettro}.} \label{tab:randomS}
\begin{tabular}{lccc}
\hline
Parameter & Symbol & Our Source\\
\hline
Braking index & $n$ & 3 \\
Initial spin-down age (yr) & $\tau_0$ & $19971$ \\
Initial spin-down luminosity (erg s$^{-1}$) & $L_0$ & $2.26\times 10^{36}$ \\
SNR ejected mass (M$_\odot$) & $\Mej$ & $18.60$ \\
%Far infrared temperature (K) & $T\rs{fir}$ & $38.81$ \\
%Far infrared energy density (eV cm$^{-3}$) & $w\rs{fir}$ & $0.44$ \\
%Near infrared temperature (K) & $T\rs{nir}$ & $3390.32$ \\
%Near infrared energy density (eV cm$^{-3}$) & $w\rs{nir}$ & $0.67$ \\
Energy break (TeV) & $E_b$ & $0.23$ \\
Low energy index & $\alpha_1$ & $1.25$ \\
High energy index & $\alpha_2$ & $2.49$ \\
%Containment factor & $\epsilon$ & $0.23$ \\
Magnetic fraction & $\eta$ & $0.08$ \\
ISM density (cm$^{-3}$) & $n_{\rm{ISM}}$ & $0.68$ \\
\hline
\end{tabular}
\end{table}
%
%%%%%%%%%
%%%% FIGURA EVOLUZIONE SPETTRALE VARIE FASI
\begin{figure*}
\centering
\includegraphics[width=0.95\textwidth]{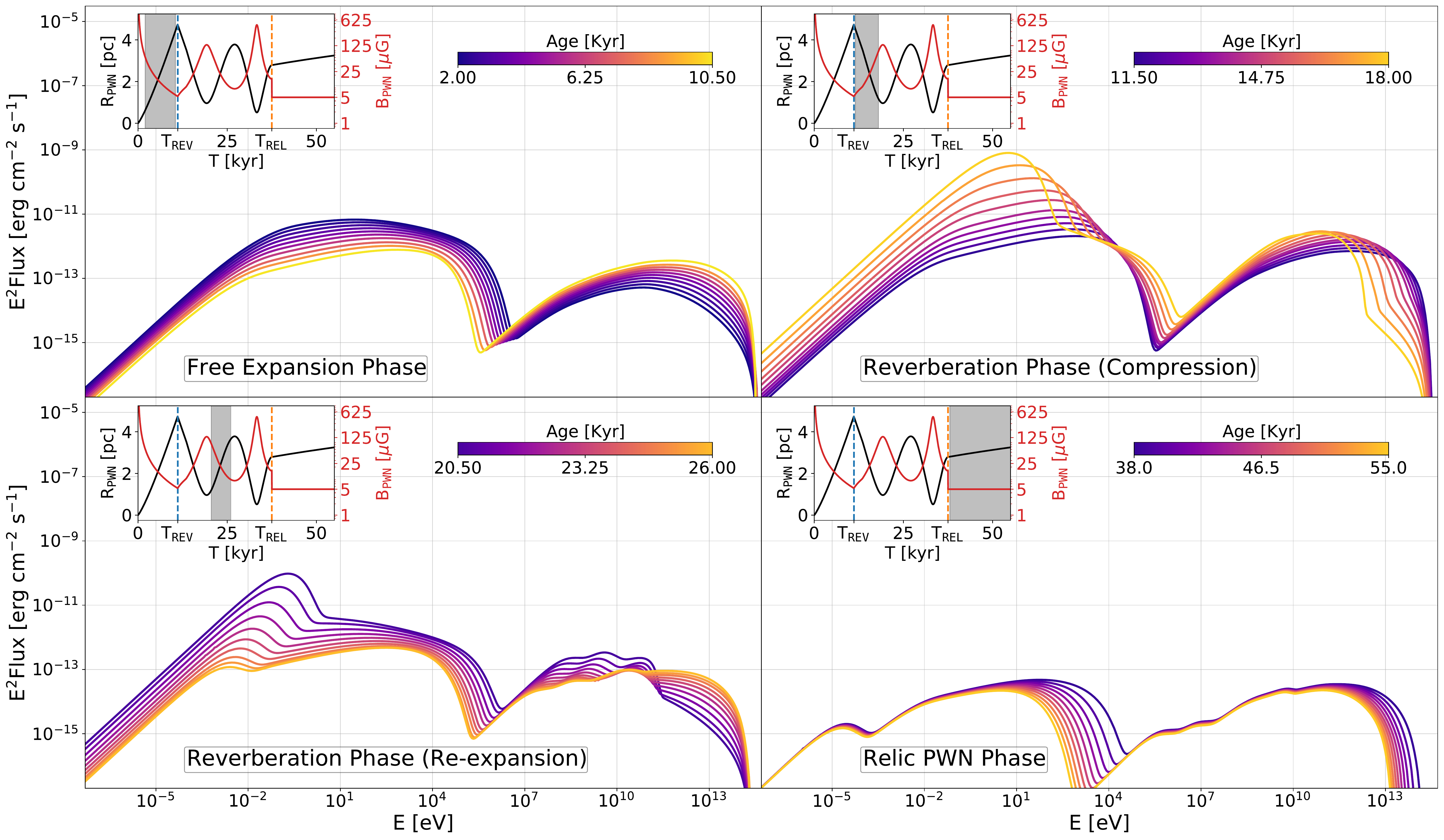}
\caption{Spectral evolution of a (randomly extracted) source in the modeled population during different evolutionary phases. 
In all panels a sub-panel illustrates the evolution of the radius (in black) and of the magnetic field (in red) with time, from 0 to the time at which the PSR eventually escapes from the SNR bubble ($T_{\rm{REL}}$). Notice that the magnetic field in the relic stage is not zero but fixed to 5 $\mu$G. The grey area in the sub-panels highlights the stage at which the spectra in each panel are extracted. The exact time to which each of the plotted spectra corresponds can be read from the color bar.}
\label{fig:spettro}
\end{figure*}
The magnetic fraction $\eta$ is taken constant in time for each PSR, randomly chosen in the range $\eta \in [0.02-0.2]$, meaning that between 80\%-98\% of the pulsar spin-down luminosity goes into accelerating particles. 
Such small values of $\eta$ are a common feature of one-zone models, but at odds with the results of 3D MHD simulations, which require a high magnetization of the wind at injection in order to produce nebular magnetic field strengths in agreement with observations \citep{Porth:2014, Olmi:2016, Olmi:2019, Olmi:2019a}. This discrepancy is due to the different evolution of the field and plasma components in the two approaches. While in 3D MHD simulations effective magnetic dissipation ensures efficient energy transfer from the field to the plasma, this effect is not included in one-zone models. In fact, in these models, due to radiation losses, the ratio between magnetic and particle energy in the nebula is always larger than at injection.
Once the magnetic and particle energy content of the PWN are known, the pressure $\Ppwn$ can be easily computed (Eq. B8 to Eq. B11 in \citealt{Fiori:2020}).
The energy distribution, and related spectral evolution of each source in our sample is computed solving numerically Eq. \ref{p_transp} coupled with the PWN dynamical evolution. To this end, we adopt {\sc GAMERA} \citep{Hahn:2015}, a freely available {\sc C++} library, with a {\sc Python} wrapper, that allows to compute the spectra of a large number of high-energy astrophysical sources.

In Fig.~\ref{fig:spettro} we show the expected photon spectral evolution for one of our sources during each of the characteristic phases that have been previously described in Sect.~\ref{sec:pwne_dyn}. The evolution of the PWN radius and magnetic field strength are also shown. The corresponding input parameters are listed in Table \ref{tab:randomS}.
During the free expansion phase, the synchrotron emission, extending from the radio band up to a few tens (or even hundreds for the brightest sources) of MeV, decreases with time. The MeV cutoff moves to lower energy, due to the decreasing magnetic field. Similarly, the break between radio and optical wavelengths, which reflects the break in the injected particle population, moves to lower energy. 
Due to the fading of the magnetic field during this phase, the synchrotron cooling break moves to higher frequencies (being $B\propto t^{-p}$ with $p\geq 1$ and $\nu_B \propto B^{-3} t^{-2}\propto t^{3p-2}$).
In this model, characterized by a large $\tau_0$ ($\tau_0 > \tch$), so that particles are efficiently injected during all the free expansion phase, this implies that particles injected at later times are less affected by synchrotron losses. 
This causes a progressive hardening of the of the X-ray spectrum during this phase, and also the ICS emission to peak at increasingly higher energies.
The change of the main target radiation from the CMB to the IR-optical, which has a larger energy density, also leads to an overall enhancement of the ICS emission. 
The observed decrease with time of the maximum frequency at which synchrotron radiation is emitted reflects the fact that in this particular system acceleration is never radiation limited and the maximum energy is set to the pulsar potential drop.
Once the system enters the first reverberation phase, and starts experiencing compression, the trend of synchrotron emission is reversed: the MeV cutoff and the radio-optical break move to higher energies; the total synchrotron luminosity increases, and now the combination of synchrotron cooling and adiabatic gains leads to a very steep spectrum in the optical to X-ray range \citep{Bucciantini_Arons+11a}.
The ICS emission keeps rising, but now it peaks at progressively lower energies, due to the burn-off of the electrons able to interact with more energetic photons than the CMB ones.
After the compression phase, the system experiences a re-expansion. The various spectral features of the synchrotron portion of the spectrum change in a similar way to the initial phase of free-expansion. 
The latest phases of evolution show a more structured ICS spectrum, with a shape that depends on the properties of the synchrotron spectrum at the moment the PWN enters the reverberation process.

The system shown in Fig.~\ref{fig:spettro} corresponds to a PSR that escapes from the parent SNR around 30~kyr after the core collapse SN. The relic nebula that is left inside the SNR will continue to emit synchrotron radiation in the residual ambient field, that is modelled according to Eq.~\ref{eq:Bfield2}, with the injection term
set to zero. It is interesting to note that the synchrotron portion of the spectrum arises accordingly to the evolution of the last particles injected into the system before escape. Given that neither the particle content nor the magnetic field change appreciably in this phase (diffusion is negligible for most particles) the only spectral change is related to the location of the synchrotron (and related ICS) cutoff, that decreases due to radiation cooling.

\section{Results and discussion}
\label{sec:results}
We computed the TeV emission of each source in our sample (around 1300 objects), considering as the emitting area the entire region within the PWN radius or, in the case of relic systems, the radius of the radio bubble in adiabatic expansion.
%
%%%%%%%%%%%%%%%%%%%%%
\begin{figure}
\centering
\includegraphics[width=.48\textwidth]{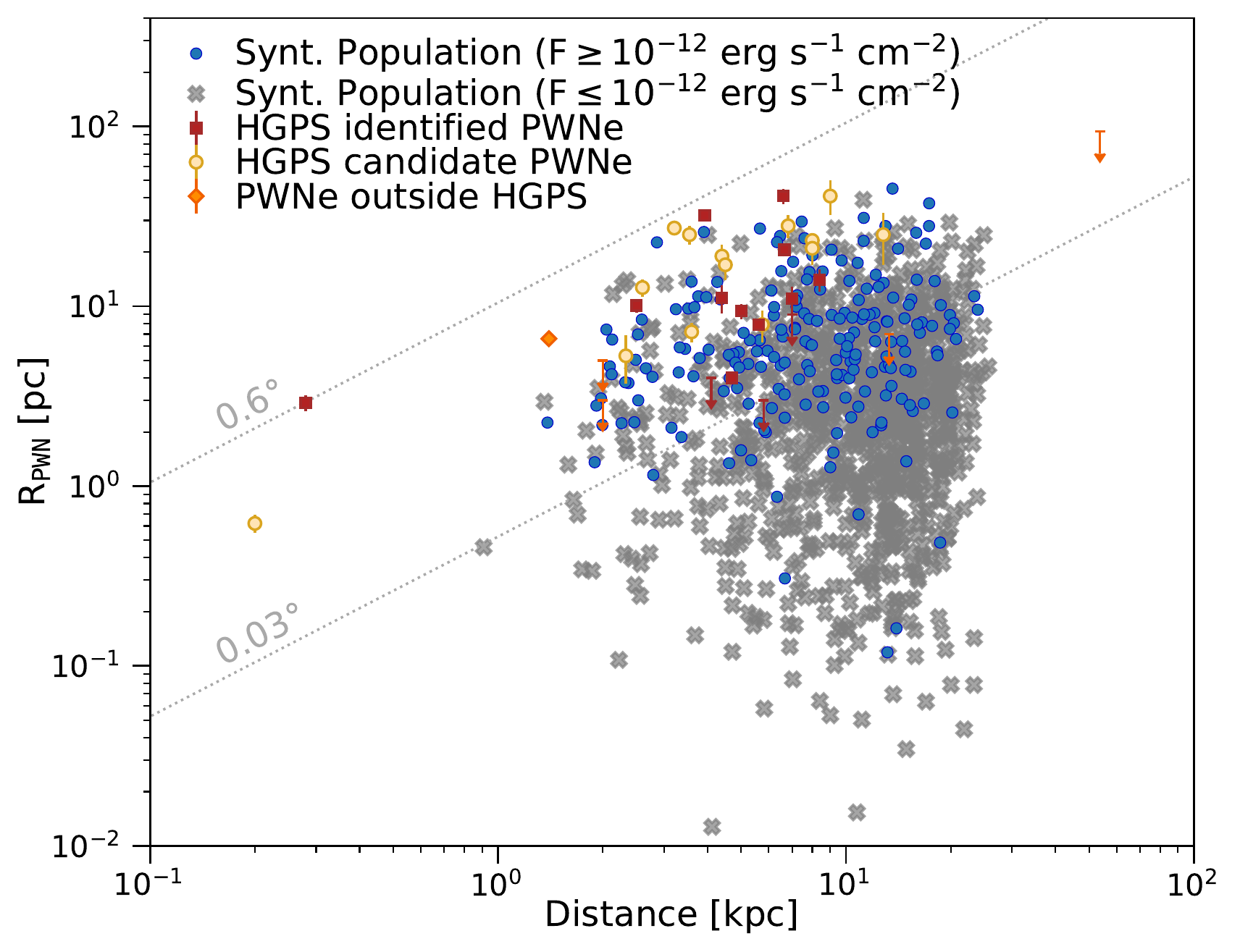}
 \caption{The TeV sizes of our synthetic population of PWNe and of those discussed within the HGPS context are reported, versus distance of the system. For the synthetic population the system extension is taken to coincide with the nebular radius at given age. Different symbols are for different sub-classes as specified in the plot legend. The flux threshold at $F=10^{-12}$erg s$^{-1}$ cm$^{-2}$ is taken from \citealt{HESScoll:2018-PWN}. Dotted lines refer to the minimum ($0.03^\circ$) and maximum ($0.6^\circ$) angular extension estimated by \citealt{HESScoll:2018-GPS} from PWNe detected in the HGPS.}
 \label{fig:plotR}
\end{figure}
%%%%%%%%%%%%%%%%%%%%%
\begin{figure*}
\centering
\includegraphics[width=0.47\textwidth]{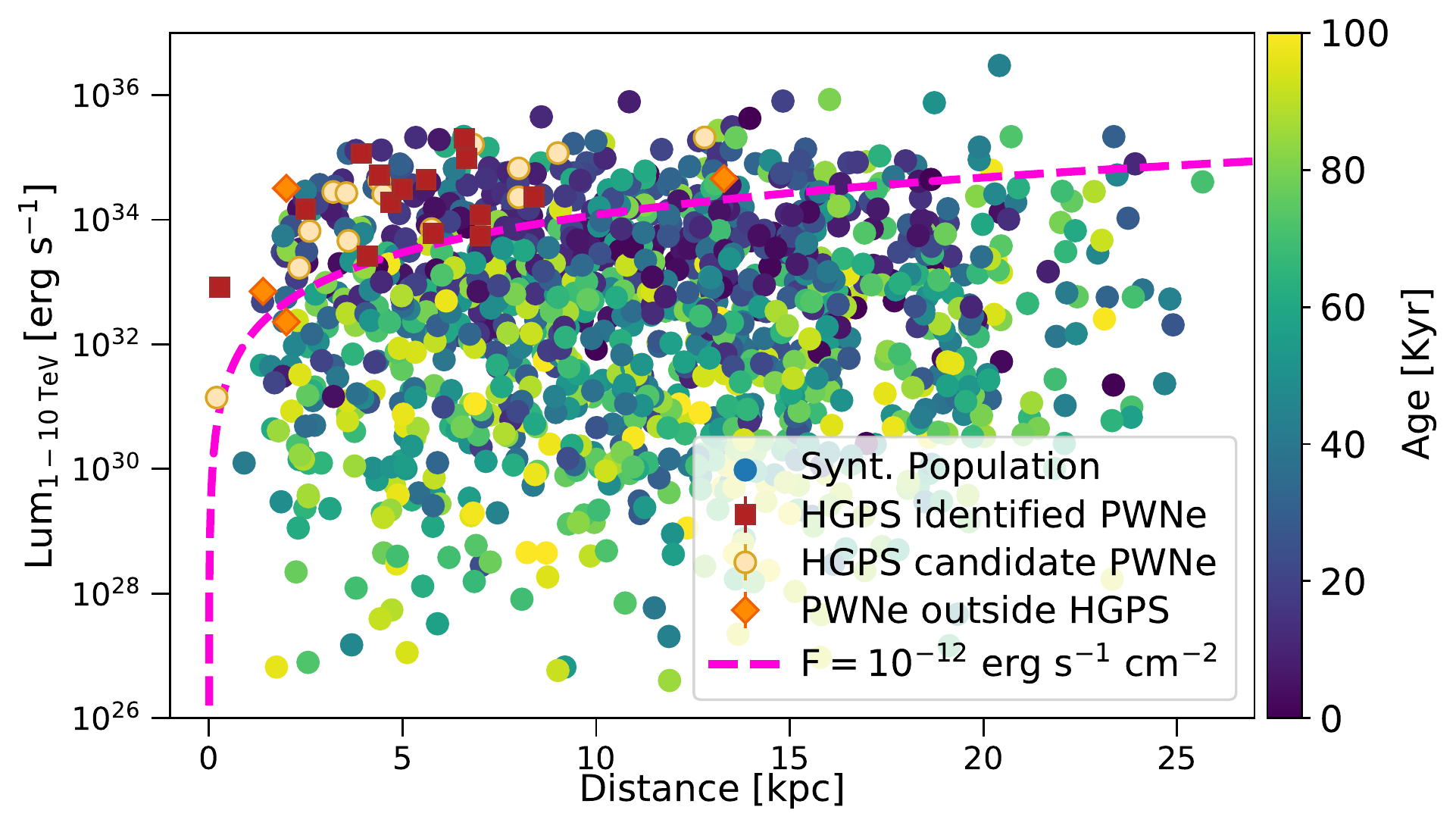}\,
\includegraphics[width=0.48\textwidth]{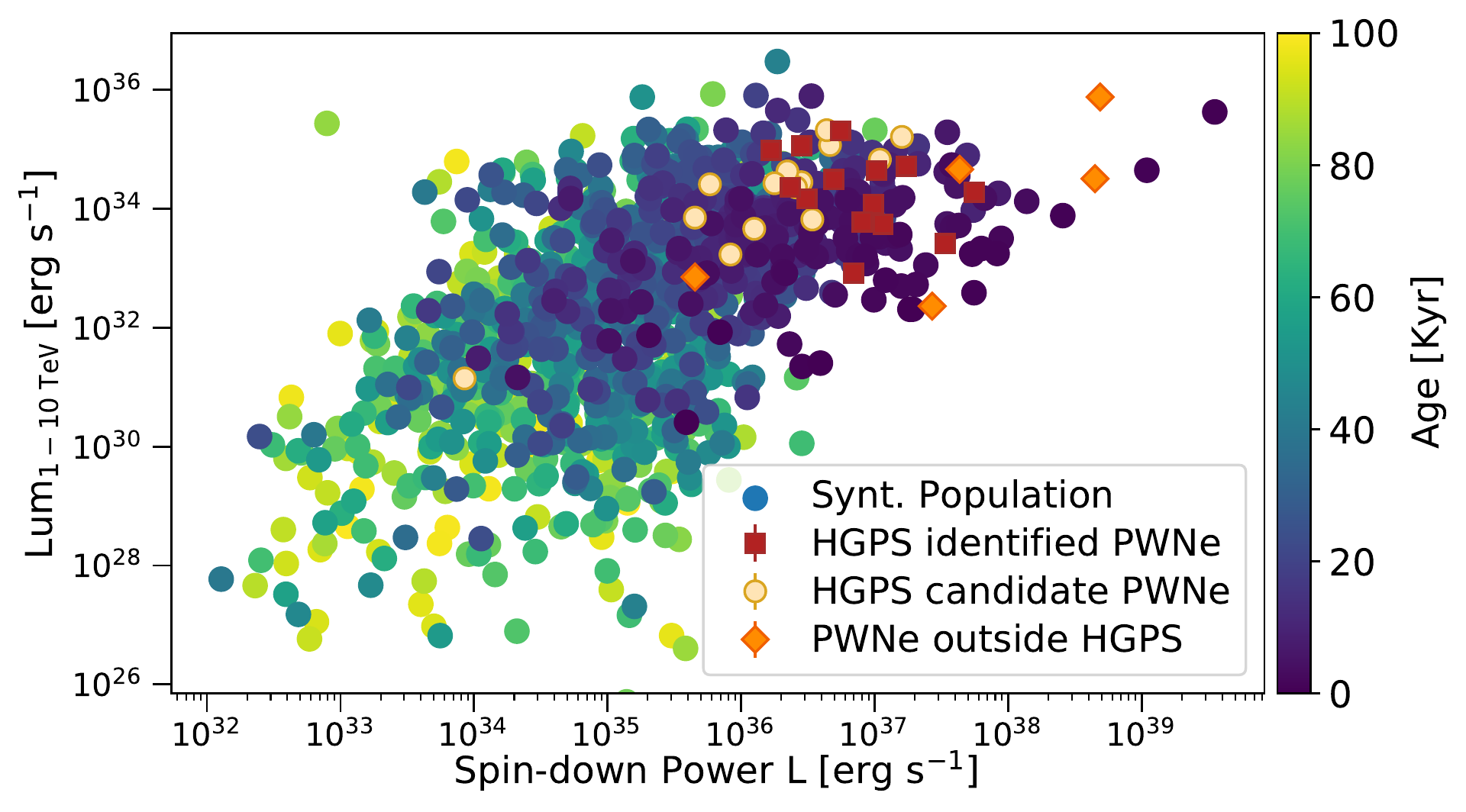}\\
\includegraphics[width=0.48\textwidth]{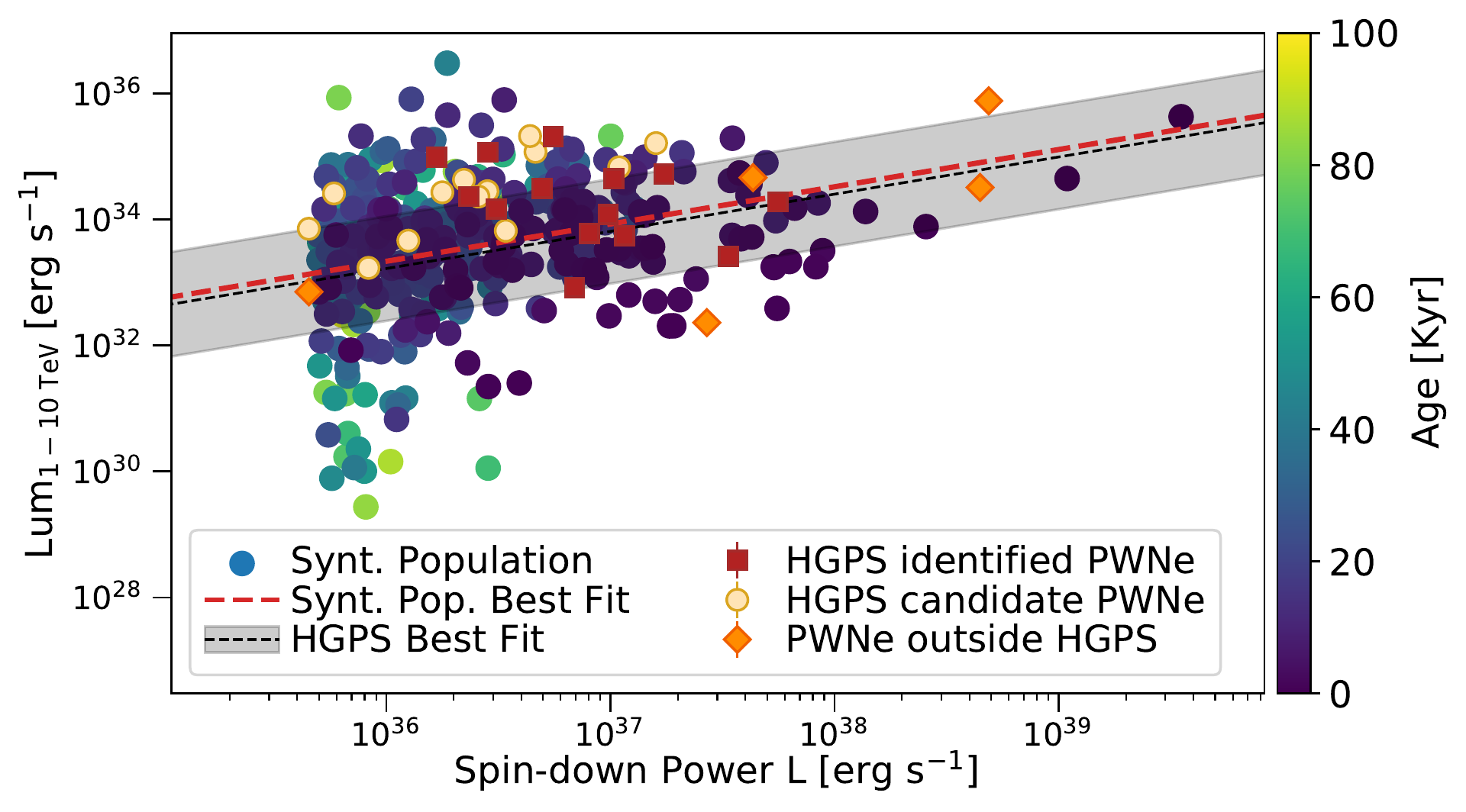}\,
\includegraphics[width=0.48\textwidth]{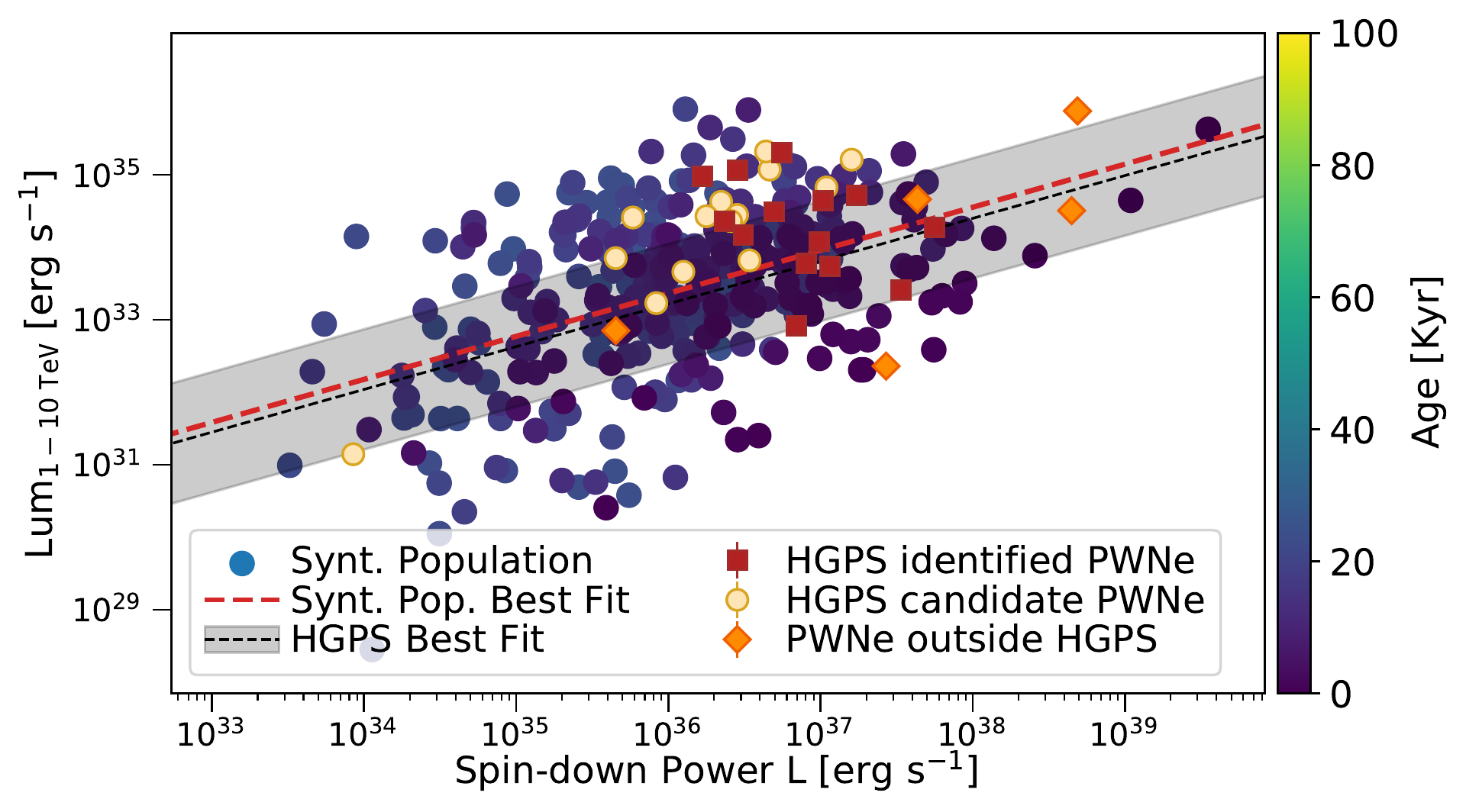}\\
\caption{\textit{Upper panels:} \g-ray luminosity in the 1-10 TeV band ($L_{\rm 1-10\ TeV}$) versus source distance (left panel) and pulsar spin-down power (right panel). The entire synthetic population is represented together with the HGPS PWNe (the symbols are specified in the plot legend). The dashed pink curve in the upper-left panel is the H.~E.~S.~S. detection threshold flux.
\textit{Bottom panels:} PWNe representation in the $L_{\rm 1 - 10\ TeV}$- $\dot E$ for two different selections of the synthetic population, namely: sources with $\dot{E}$ > $5\times10^{35}$ erg $s^{-1}$ (left panel);  sources with age $<25$ kyr (right panel). 
In both cases we show the best fitting relation between the \g-ray luminosity and spin-down power as found in the H.~E.~S.~S. data (black dashed line with standard deviation represented as a shaded grey area) and as determined on our selected population (red dashed line). In both cases we found an excellent agreement with the best fit to the  H.~E.~S.~S. data.}

\label{fig:TeVplots}
\end{figure*}
In Fig.~\ref{fig:plotR} we show the distribution of PWNe radii versus the source distance, with the indication of the maximum and minimum extensions found in the HGPS ($0.6^\circ$ and $0.03^\circ$ respectively).
We plot the simulated PWNe in blue (gray) color if the system is above  (below) the H.~E.~S.~S. detection threshold flux of $F=10^{-12}$ erg s$^{-1}$ cm$^{-2}$  \citep{HESScoll:2018-PWN}.
Despite the simplified description of the nebular geometry used in our model, and the lack of asymmetric systems in the population, we found a remarkable agreement with data in the distribution of extensions at TeV.

%%%%%%%%%%%%%%%%%%%%%
In Fig.~\ref{fig:TeVplots} we represent our population in the $L_{\rm 1-10\ TeV}$-Distance (upper left panel) and $L_{\rm 1-10\ TeV}$-$\dot E$ space (all other panels), with $L_{\rm 1-10\ TeV}$ the PWN luminosity in the 1-10 TeV range.
The upper left panel collects the entire population of synthetic plus HGPS PWNe. The magenta curve represents the H.~E.~S.~S. average detection threshold, corresponding to $3\times 10^{33}$ erg s$^{-1}$ for a source at 5.1 kpc distance.
As one would expect, the vast majority of the synthetic population lies below the H.~E.~S.~S. detection threshold: only $\sim10\%$ of the systems with $80\,\mathrm{kyr}\,\leq t_{\mathrm{age}} \leq 100\;\mathrm{kyr}$ are above the detection limit.
The present firmly identified PWNe can all be found within $\sim 15$ kpc and are relatively young, with ages $\lesssim 40$ kyr.
A much larger fraction of the population is expected to be revealed with the advent of CTA, whose anticipated detection threshold is more than one order of magnitude lower than the H.~E.~S.~S. one \citep{Remy:2021}.

In the other three panels of  Fig.~\ref{fig:TeVplots}, we show the TeV luminosity as function of the pulsar spin-down power.
Again, we notice that the detected sources only represent the most powerful ($L(t)\gtrsim 5\times 10^{35}$ erg s$^{-1}$) and TeV brightest ones of the entire population.
The possible existence of a correlation between the TeV luminosity and the pulsar spin-down power, in analogy to what observed in the X-rays, was investigated by \citet{Kargaltsev:2013}, who found no clear evidence for such a correlation.
On the other hand, the analysis of the wider {\it Fermi} energy band (0.1 GeV -- 100 GeV), including also less luminous and less energetic systems than those considered by \citet{Kargaltsev:2013}, indicates a linear dependence of the \g-ray luminosity on the spin-down power: $L_\gamma\propto \dot{E}$ \citep{Abdo2013}.
More recently, the analysis of the most updated H.~E.~S.~S. data \citep{HESScoll:2018-PWN} suggested indeed a possible correlation between the TeV luminosity and the pulsar power, but with a flatter dependence than in the {\it Fermi} band: $L_{1-10\ \rm{TeV}}\propto \dot{E}^{(0.59\pm0.21)}$.
From the analysis of our population, we surprisingly find a correlation more similar to what deduced from  the {\it Fermi} analysis than from the H.~E.~S.~S. one. 
We in fact obtain a slightly super-linear trend: $L_{1-10\ \rm{TeV}}\sim\dot{E}^{(1.13\pm 0.06)}$.
This might indicate that the submerged part of the non-detected PWNe, actually dominates the relation between the emitted \g-ray luminosity and the pulsar spin-down power.
It is interesting to notice that a trend similar to that found in the H.~E.~S.~S. data can be retrieved with an appropriate selection of the population. 
If we only consider sources with $\dot{E} > 5 \times 10^{35}$ erg s$^{-1}$, we get  $L_{1-10\rm{TeV}}\sim\dot{E}^{(0. 59\pm 0.11)}$. In the same way, considering only systems younger than 25 kyr, we find  $L_{1-10\rm{TeV}}\sim\dot{E}^{(0.59\pm 0.06)}$. %
The direct comparison of these two selections with the trend extracted from the H.~E.~S.~S. data is shown in the lower panels of the same Fig.~\ref{fig:TeVplots}.
This exercise clearly shows that the sensitivity of the observations introduces a bias in determining the $L_{1-10\ {\rm TeV}}$-$\dot E$ relation. At the same time, it shows that our PWNe population well represents the observed ones in the $L_{1-10\ {\rm TeV}}$-$\dot E$ plane, as long as proper cuts are adopted.

\begin{figure*}	
\centering
	\includegraphics[width=.48\textwidth]{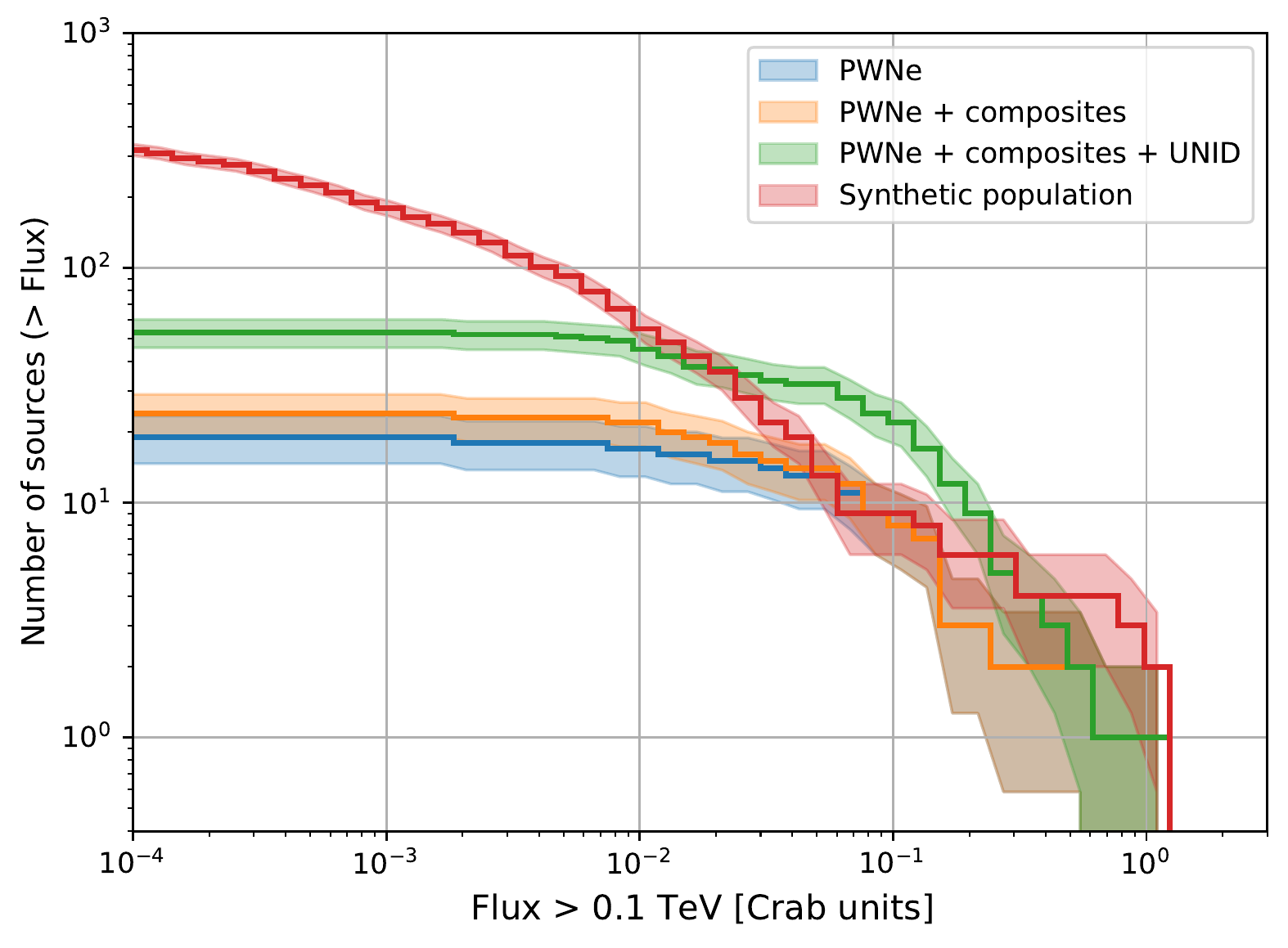}     
	\includegraphics[width=.48\textwidth]{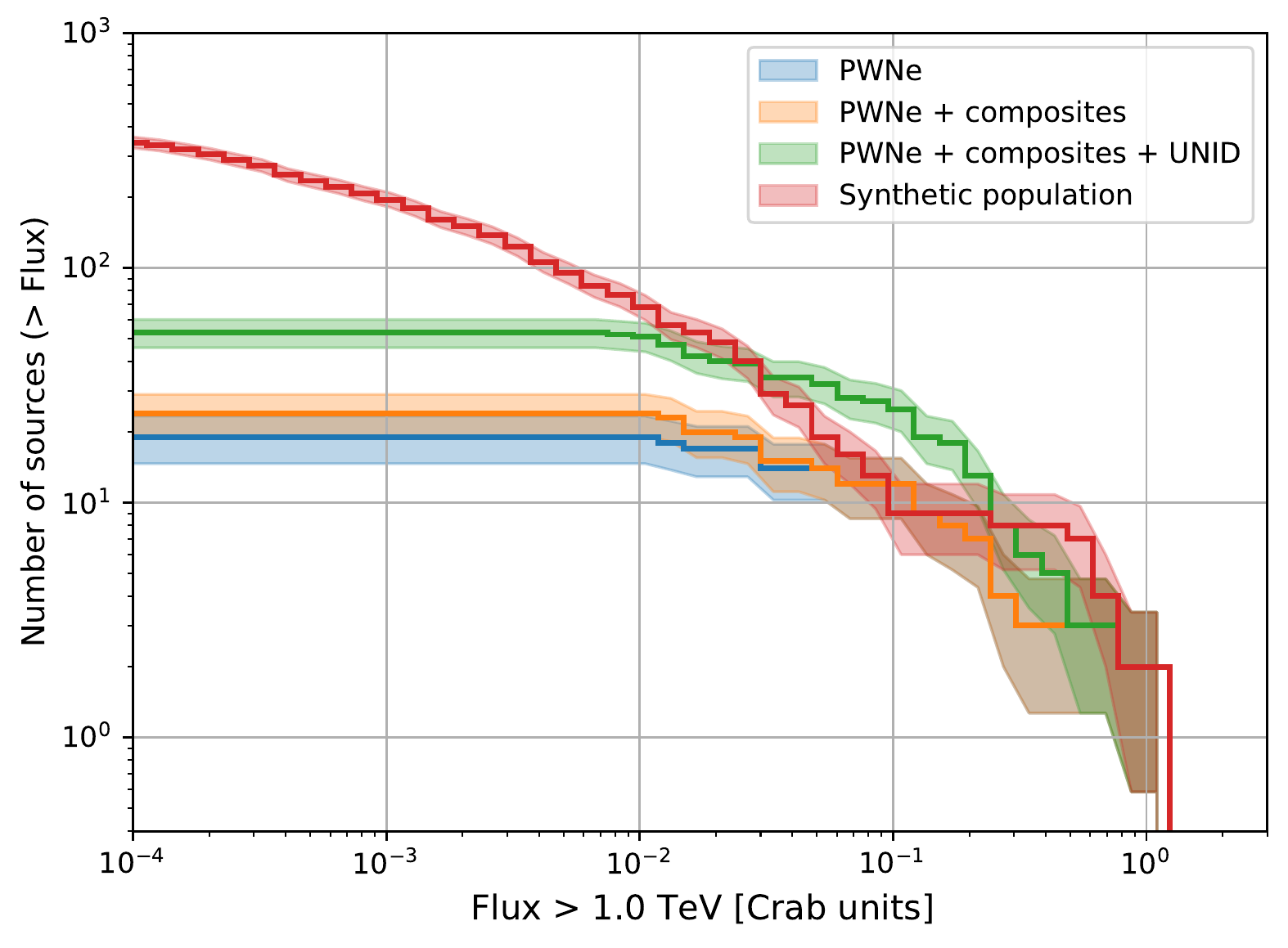} 
    \caption{Logarithmic plot of the number of sources emitting in a specific flux range:  $>0.1$ TeV (left panel) and $>1$ TeV (right panel). The synthetic population (in red) is directly compared with the firmly identified PWNe (in blue), PWNe in a composite SNR (in orange) and the sum of these two with the known unidentified sources (in green). The lighter colored areas represent the  errors of each curve.}
    \label{fig:NS}
\end{figure*}
%%%%%%%%%%%%%%%%%%%%%
In Fig.~\ref{fig:NS}, we compare the total number of sources with \g-ray flux above 0.1 TeV (plot on the left) and 1.0 TeV (plot on the right), with fluxes expressed in Crab units (CU).
To compare with available observations, we limit our analysis to the Galactic plane, with  $|\mathrm{GLAT}|<2^\circ$ and $\mathrm{GLON}<70^\circ \cup \mathrm{GLON}>270^\circ$. 
The synthetic population perfectly reproduces the PWNe (and PWNe +  composite) data down to $10^{-2}$ CU in both cases. Beyond that value we see that the observational curves flatten due to the loss of instrumental sensitivity, and below that flux any comparison is meaningless.
We also notice that the synthetic population accounts very well for the unidentified sources with higher fluxes, possibly meaning that some of them are in fact unidentified PWNe, as one might also guess from considerations related to the total number of expected PWNe in the Galaxy.
In contrast, the unidentified population is not reproduced (by a factor of $\sim 2$) in the range $2\times 10^{-2}-10^{-1}$ CU by the synthetic PWNe population.
This might be an indication of a different nature of the sources contributing in this energy range (TeV halos?), or alternatively a problem with our model, due to the oversimplifications introduced.
%

%%%%%%%%%%%%%%%%%%%%%%%%%%%%%%%%%%%%%%%%%%%%%%%%%

We have evaluated the impact on the total TeV flux of our description of relic systems, with particular reference to the assumption that, once escaped, pulsars can never re-enter the PWN bubble, no matter the value of $V_{\rm PSR}$. 
These systems represent $\sim 13\%$ of the overall PWN population at 100 kyr. To assess how their modeling affects our results, we evaluated how the expected TeV flux changes by subtracting from our PWN population all relics whose parent pulsar had escaped before $t^{\prime}$; computing the average TeV spectrum of the remaining population; replacing the contribution of all subtracted relics with this average spectrum.
For the time $t^{\prime}$ we assumed 100 kyr (corresponding to replacing all relics), 70 kyr and 50 kyr. We found corresponding flux increases by 11\%, 6\% and 4\%.
%
%
%%%%%%%%%%%%%
\begin{figure*}
\centering
	\includegraphics[width=.45\textwidth]{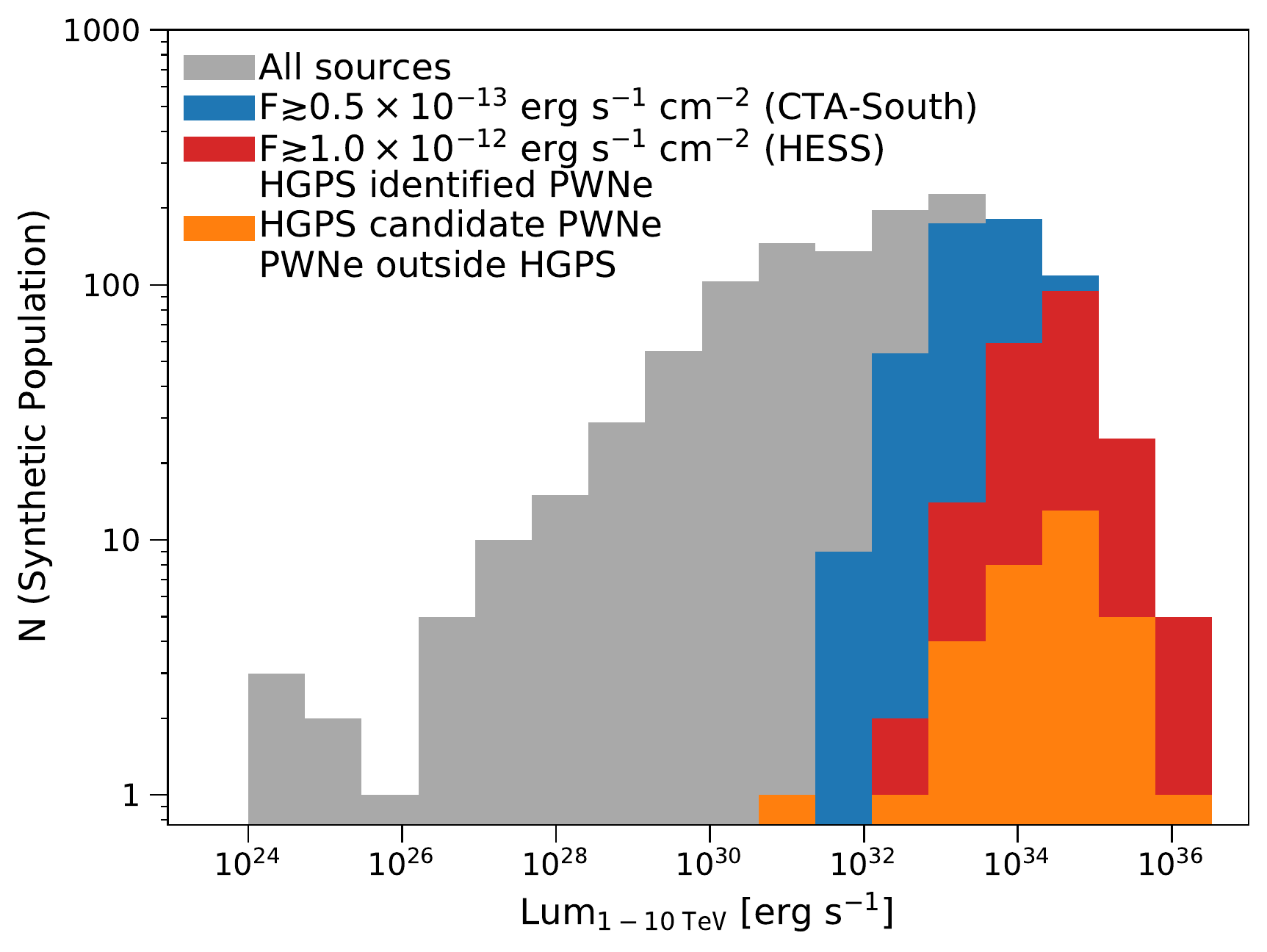}\\
	\includegraphics[width=.45\textwidth]{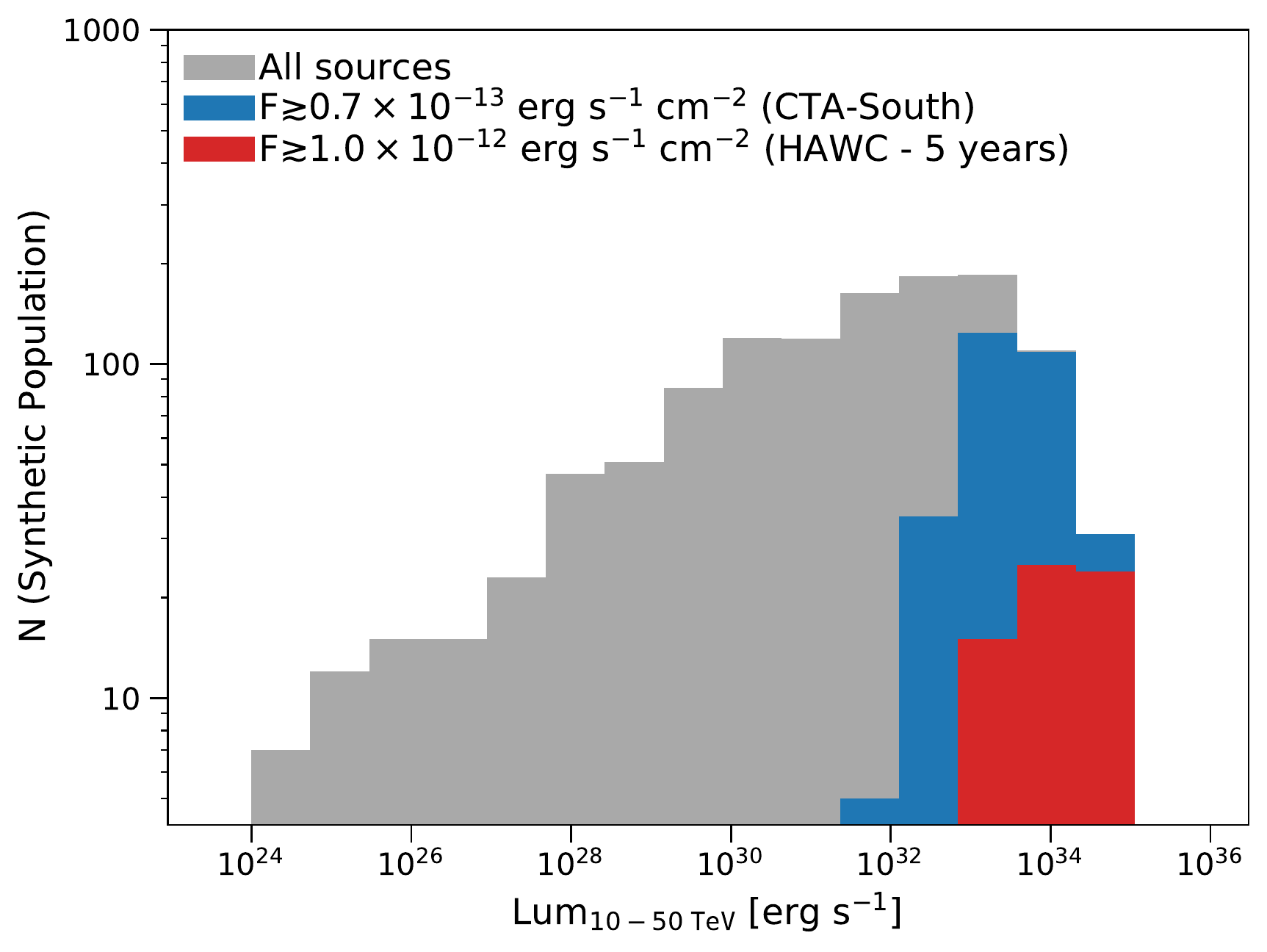}\,
	\includegraphics[width=.45\textwidth]{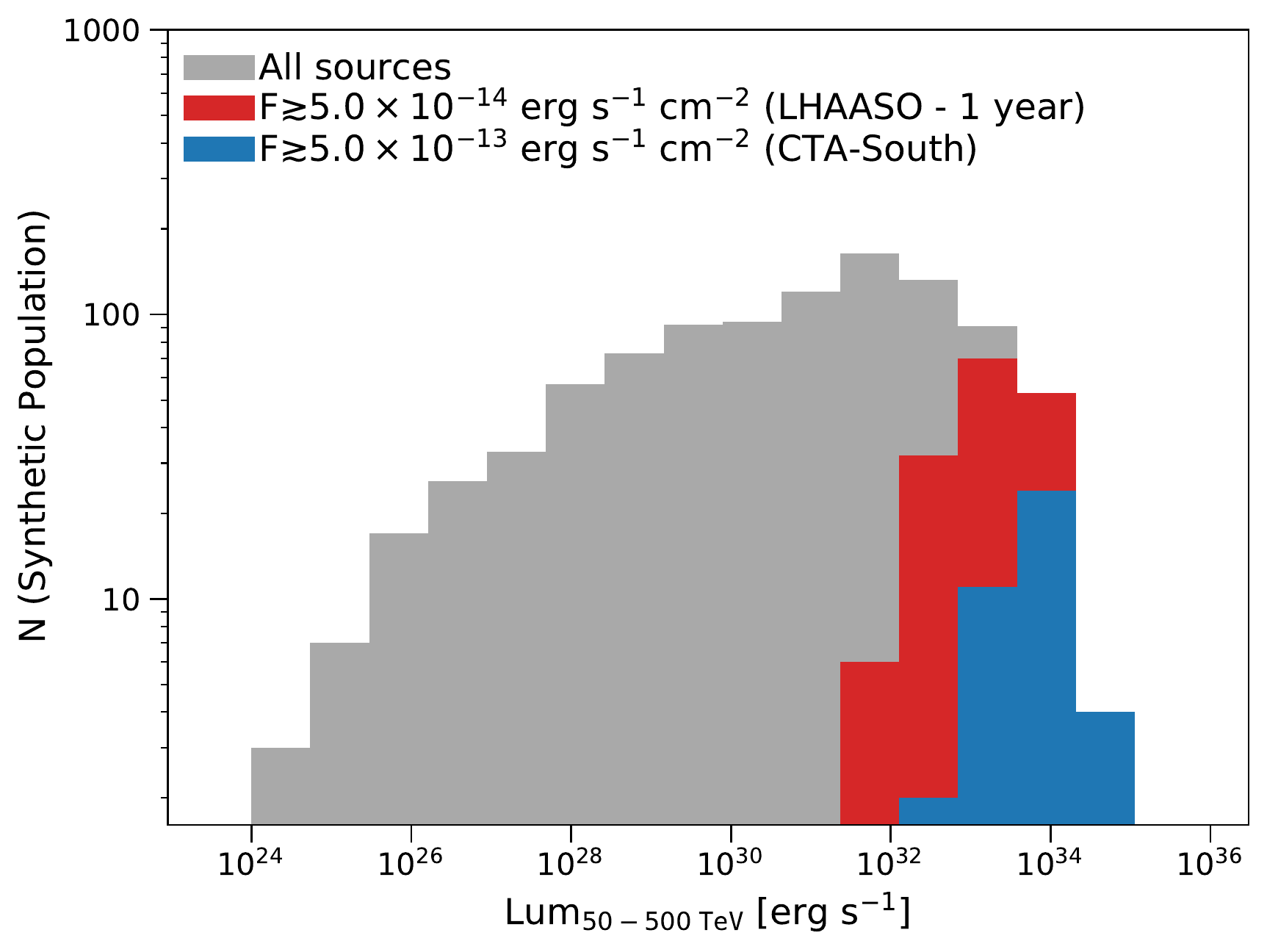}
    \caption{Total number of synthetic PWNe per luminosity beam in different \g-ray luminosity ranges (grey) and number of detectable sources by current and upcoming $\gamma$-ray instruments.
    {\it Upper panel:} sources above the H.~E.~S.~S. threshold (red) are compared with the actual detected ones (orange) and with those above the estimated CTA threshold (from \citealt{Remy:2021}, blue) for the 1-10 TeV range.
    {\it Bottom left panel:} expected detections by HAWC (red) and CTA (blue) in the 10-50 TeV range, with detection thresholds estimated from \citet{HAWCsens} and \citet{Remy:2021}, respectively.
    {\it Bottom right panel:} LHAASO (red) vs CTA (blue) in the 50-500 TeV range, with the LHAASO detection threshold estimated from \citet{LHAASOsens}.}
    \label{fig:NsourceG+ALL}     
\end{figure*}
%%%%%%%%%%%%%%%%%%%%%
%
This means that, even if relic systems are poorly described from the dynamical point of view, their contribution is such that the global \g-ray emission is only affected by a maximum error of a few percent.

\begin{figure*}	
\centering
	\includegraphics[width=.495\textwidth]{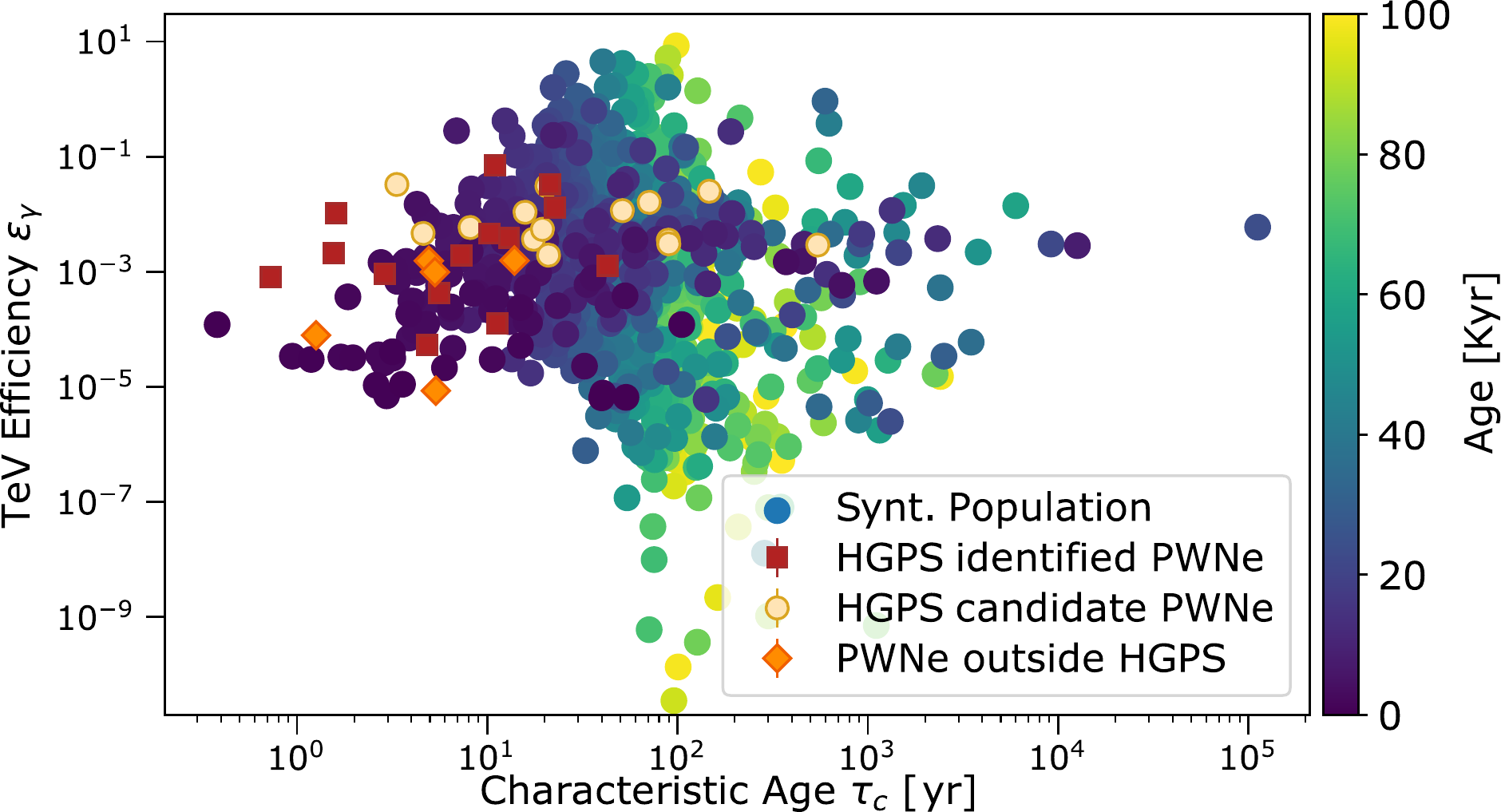}	
	\includegraphics[width=.495\textwidth]{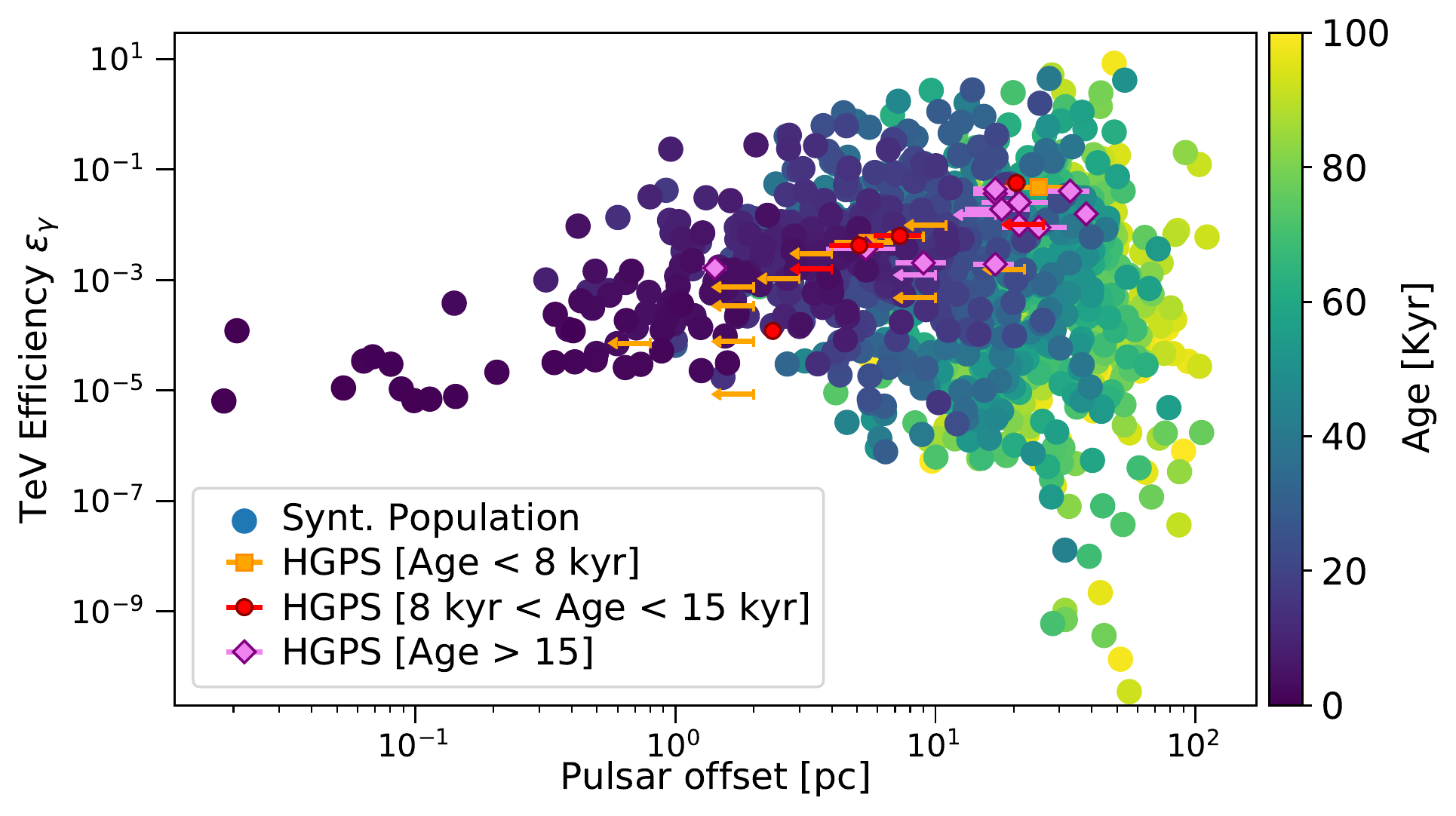}
    \caption{\textit{Left panel}: Evolution of the TeV efficiency $\epsilon_\gamma=L_\gamma/\dot E$ with the pulsar characteristic age $\tau_c$, considering the 1-10 TeV range for the \g-ray luminosity.
    \textit{Right panel}: Variation of the \g-ray efficiency as a function of the pulsar offset for different age groups (shown in different colors).
    }
    \label{fig:effg}
\end{figure*}

In Fig.~\ref{fig:effg} we show the \g-ray efficiency in the 1-10 TeV energy range, $\epsilon_\gamma=L_\gamma/\dot{E}$, as a function of the characteristic pulsar age $\tau_c$ (panel on the left) and of the pulsar displacement (panel on the right), comparing  with the same quantities as obtained from observations. Information on the source ages is also included with a color-code.
Conversely to X-ray efficiency (namely $L_X/\dot{E}$), that traces the most recent evolution of the source, the \g-ray efficiency traces the emission along the history of the source, thus values larger than unity are not unexpected.

On the other hand, those systems would be the oldest, with the largest displacements and extensions. As a result we can expect that a sizeable fraction of the high \g-ray efficiency systems would not be detected.
Having access to the entire population without observational biases, here we can confirm that higher efficiencies at \g-rays are characteristic of objects older than 8 kyr; an efficiency higher than unity is only found for ages $>25$ kyr and mostly for large pulsar offsets. 
The occurrence of systems with $\epsilon_\gamma\gtrsim1$ can be easily interpreted recalling that $\epsilon_\gamma$ is the ratio between the current values of \g-ray luminosity and pulsar spin-down power, but the former is the result of the entire injection history (as already pointed out by \citealt{HESScoll:2018-PWN}). 
In any case, we see that only a relatively small number
of evolved systems show an efficiency $\epsilon_\gamma\gtrsim0.1$.

%%%%%%%%%%%%%%%%%%%%%
%%%%%%%%%%%%%%%%%%%%%

In Fig.~\ref{fig:NsourceG+ALL} we show the PWN distribution as a function of luminosity in different energy ranges, and compare our synthetic population (grey) with the number of sources above the detection threshold of existing and upcoming facilities.
In the upper panel we consider sources with luminosities in the range 1-10 TeV and report both the sources that H.~E.~S.~S. can detect (red) and those detected (orange, including also PWN candidates and those marked as outside HGPS), in addition to  sources (blue) above the CTA threshold \citep{Remy:2021}.
One can see that a very large number of sources is expected at $L_\gamma<10^{32}$ erg s$^{-1}$, and tens of undetected sources with luminosity above the H.~E.~S.~S. threshold are also expected. The reason for this can be understood from comparison with Fig.~\ref{fig:TeVplots}, from which we see that most of these will be at distances >15 kpc, older than 40 kyr and characterized by a residual spin-down luminosity lower than $10^{35}$ erg s$^{-1}$, which makes their identification in the TeV sky extremely challenging. 
This discrepancy will possibly much reduce with the advent of CTA, for which the number of sources above threshold increases by a factor of $\sim 3$ (560 vs 200).

In the lower panels of Fig.~\ref{fig:NsourceG+ALL} we compare our synthetic population in the 10-50 TeV range (left) with the expected sources detectable by HAWC (red) and CTA (blue), and in the 50-500 TeV range (right) with LHAASO (red) and CTA (blue).

Finally, we think it appropriate to briefly discuss the distribution of sources in the different evolutionary phases at the end of our simulation, and compare the results with available data.
To date we know about 20 sources in the free-expansion phase (see e.g. \citealt{Torres:2014}, \citealt{Zhu2018}), while only a very limited number of objects have been confirmed to be in a reverberation stage, namely showing direct evidence of the interaction between the PWN and the SNR reverse shock, as observed e.g. in Vela X  \citep{Blondin:2001}, Boomerang \citep{Kothes_Reich+06a} and the Snail \citep{Ma_Ng+16a}).

In our synthetic population, at the end of the simulated time, we find the sources distributed as follows:
\begin{itemize}
    \item $\sim 11\%$ of the population (132 sources) is in the free-expansion phase; among these 103 sources are above the H.~E.~S.~S. threshold flux (27 in the visibility range $L>5\times 10^{35}$ erg s$^{-1}$; $F>10^{-12}$ erg s$^{-1}$ cm$^{-2}$ discussed in Fig.~\ref{fig:TeVplots} -- about $2.2\%$ of the population); 
    \item $\sim 76\%$ of the sources have entered the reverberation phase (948 sources); among these 236 are in the H.~E.~S.~S. detectability range (94 in the visibility region -- $\sim 7.5\%$ of total);
    \item $\sim 13\%$ of the sources are relic (174 sources); only 4 of these are above the detection threshold (3 in the visibility range -- $\sim 0.2\%$ of total).
\end{itemize}
As expected, the majority of the sources are middle-aged, and hence are in - or have passed through - the reverberation phase. How these numbers compare with the catalogue of observed TeV sources will be discussed in the following Section.

%%%%%%%%%%%%%%%%%%%%%%%%%%%%%%%%%%%%%%%%%%%%%%%%%%
\section{Conclusions}
\label{sec:conclusions}
\label{lastpage}
Being PWNe the most numerous sources expected to be detected in future \g-ray observations, the problem of how to correctly account for their contribution to the overall \g-ray emission is extremely topical.
In this paper we proposed a physically motivated model of the Galactic PWNe population responsible for the \g-ray emission, taking into account the available observational constraints from known sources.
A noticeable difference with respect to previous results is in the pulsar population that we have assumed: we found that in fact the PWN population in the Galaxy is best reproduced when assuming the properties of the powering pulsars as deduced from the \g-ray emitting pulsar population, rather than from the entire radio pulsar population. This is due to the fact that \g-ray pulsars are representative of a younger population, including the only objects that can indeed power PWNe.
The PWNe population was constructed by associating each pulsar of the synthetic population to a core-collapse SNR, using a Monte-Carlo method.
The entire population was then evolved for $10^5$ yr with a modified one-zone model, that incorporates an approximate but reliable recipe to properly account for, both in terms of dynamics and spectral evolution, the complex transition between the free-expansion phase and the late stages, through the so-called reverberation phase.
During reverberation each PWN might experience a sequence of compressions and re-expansions, (depending on its energetics and on the properties of the parent SNR), that can modify the spectral properties at the late stages. This phase had not received much attention in the past, since it requires in principle a complex treatment to follow the dynamical evolution of each single source.
Nonetheless, it cannot be ignored when the purpose is the modeling of late time \g-ray emission, since this will very significantly depend on the past injection history.

Here we adopted a simplified, but physically motivated model, which was proven to provide a good description of the reverberation phase, based on the results of a large number of HD simulations.
In particular our model takes care of the problem of
the artificial over-compression introduced by standard one-zone models, that impose the thin-shell approximation also during reverberation. 
We compare the properties of the evolved population with available data (mainly from the HGPS) and find a very good agreement, especially when considering the intrinsic biases introduced by observational limits.
Our simulated population counts around 200 sources above the H.~E.~S.~S. flux detection threshold of $\sim 10^{-12}$ erg s$^{-1}$ cm$^{-2}$.
These reduce to 124 when considering also the limit on the pulsar luminosity of $L\gtrsim 5\times10^{35}$ erg s$^{-1}$ discussed in Fig.~\ref{fig:TeVplots}.
To date, the second TeVcat catalogue reports a total of 38 TeV sources marked as PWNe or halos, plus $\sim 70$ unidentified sources, for a total of $\sim 110$ sources. The HGPS found 14 firmly identified PWNe plus $\sim 45$ unidentified sources, for a total of $\sim 60$ TeV sources possibly associated with PWNe.
Then around 1/3 of the synthetic population above the H.~E.~S.~S. flux detection threshold can be considered as detected, even if the largest part of the sources have not been identified.
The discrepancy between the theoretically expected sources above threshold and the number of TeV detected sources decreases if we consider the visibility limit: of the 124 sources in this region, 1/2 have been detected.
The reasons why around $\%50$ of the sources are still missed can be manifold:  lack of spatial coverage of present instruments; source confusion and ensuing difficulty of identification; too large angular extension of the sources.

The situation will change dramatically with CTA, whose
detection threshold is expected to be lower by at least one order of magnitude \citep{Remy:2021}: the number of theoretically detectable sources will increase to $\sim 560$.
Even considering the same factor of 1/3 between revealed and expected sources -- likely an underestimate, given that CTA will also have an improved spatial coverage of the sky --, this means that around 200 PWNe should be detected in the first CTA Galactic Plane Survey.
%%%%%%%%%%%%%%%%%%%%%%%%%%%%%%%%%%%%%%%%%%%%%%%
%%%%%%%%%%%%%%%%%%%%%%%%%%%%%%%%%%%%%%%%%%%%
\section*{Acknowledgements}
%%%%%%%%%%%%%%%%%%%%%%%%%%%%%%%%%%%%%%%%%%%%
%
E.A, N.B., R.B., B.O. and L.Z. acknowledge financial support from the Italian Space Agency (ASI) and National Institute for Astrophysics (INAF) under the agreements ASI-INAF n.2017-14-H.0  and from INAF under grants "PRIN SKA-CTA" and “Sostegno alla ricerca scientifica main streams dell’INAF”,  Presidential Decree 43/2018. E.A, R.B., B.O. also acknowledge support from the INAF grant “PRIN-INAF 2019”, L.Z. from the ASI/INAF grant I/037/12/0.
This research made use of gamma-cat, an open data collection and source catalog for $\gamma$-ray astronomy. This research made use also of the following PYTHON packages: MATPLOTLIB \citep{matplotlib}, NUMPY \citep{numpy}, and ASTROPY \citep{astropy}.

\section*{Data Availability}

The data underlying this article will be shared on reasonable request to the corresponding author.
%%%%%%%%%%%%%%%%%%%%%%%%%%%%%%%%%%%%%%%%%%%%%%%%%%
% References
%%%%%%%%%%%%%%%%%%%%%%%%%%%%%%%%%%%%%%%%%%%%%%%%%%
\appendix

%%%% Appendix A %%%%%%%%%%%%%%%%%%%%%%%%%%%%%%%%%%%%%
\section{Transition from free-expansion to reverberation phase}
\label{app:A}
In this appendix we give analytic formulae that well approximate the dynamical evolution as computed for the free-expansion phase. The analytic approximations for the radius of the PWN (equal to that of the shell) and the reverse shock are respectively:
%%%
\begin{eqnarray}
\frac{R\rs{PWN}}{\Rch}&=&1.911\,(L^*\tau^*)^{1/5}\tau^*\frac{\left(1 + 0.965(t^*/\tau^*)^{0.719}\right)^{1.390}}{\left(1 + 1.157(t^*/\tau^*)^{-0.730}\right)^{1.645}};\qquad \\
\frac{R\rs{RS}}{\Rch}\;\;&=&2.253\,t^*-3.438(t^*)^2+3.198(t^*)^3-1.830(t^*)^4   \nonumber\\
&&\qquad+0.555(t^*)^5-0.069(t^*)^6\,,
\end{eqnarray}
%%%
where $t^*\equiv t/\tch$, while $\tau^*$ and $L^*$ are defined by Eq.~\ref{eq:chart}-\ref{eq:char}.
From the above formulae one can derive the time ($\tbegrev$) at which the reverberation phase begins. It is simply obtained as the time at which the curves of the PWN and the RS radii intersect.
In addition, analytic approximations for the swept-up mass and the PWN pressure, during the pre-reverberation phase, are:
%%%

\begin{eqnarray}
\frac{M\rs{shell}}{\Mej}&=&0.990\,(L^*\tau^*)^{3/5}
\frac{\left[1 + 1.036\,(t/\tau_0)^{-0.719}\right]^{4.170}}
{\left[1 + 1.157\,(t/\tau_0)^{-0.730}\right]^{4.934}}; \\
\frac{P\rs{PWN}}{\Esn \Rch^{-3}}&=&0.143\frac{(L^*\tau^*)^{2/5}}{(\tau^*)^3}\frac{\left[1+0.476\,(t/\tau_0)^{-0.743}\right]^{3.500}}{\left[1+1.543\,(t/\tau_0)^{0.760}\right]^{5.263}}.
\end{eqnarray}
%%%
These formulae have also been used to set the initial conditions at $\tbegrev$, as required to numerically compute the following evolution in the reverberation phase (as shown in Sec.~\ref{sec:reverberation}).
The quantities to be estimated are: the mass of the shell, then taken to be constant during reverberation;
the PWN radius, velocity, and pressure at $\tbegrev$.
In this way the initial conditions for the further dynamical evolution are fully determined.

%%%% Appendix B %%%%%%%%%%%%%%%%%%%%%%%%%%%%%%%%%%%%%
\section{Input parameters for the simulation}
\label{app:B}
In Table \ref{tab:SummaryInputPar} we summarize the parameters and relative distribution used as initial condition for the simulation of the PWNe population.

\begin{table*}
\centering
\caption{Summary of the input parameters used to generate the PWNe population. Values for the ISM density and IR background photon fields are not listed here since they depend on the position of each source in the Galaxy.} \label{tab:SummaryInputPar}
\begin{tabular}{lll}
\hline \hline
Parameter & Distribution & Values\\
\hline
\multicolumn{3}{l}{\textbf{PSRs population parameters}}\\
% \hlinez
Braking index & constant & $n=3$ \\ 
Mgnetic field & log-normal & $\langle \log_{10}(B/G)\rangle = 12.65$; $\,\,\sigma_{\log_{10}B} = 0.55$\\
Initial spin periods & Eq. \ref{eq:P0} & $\langle P_0\rangle = 50$ ms; $\,\,\sigma_{P_0} = 50$ ms; (truncated at 10 ms)\\
Kick velocity & double-sided exponential & $\langle v_{3D}\rangle = 380$ km s$^{-1}$ \\
\hline
\multicolumn{3}{l}{\textbf{SNRs population parameters}}\\
CC SNR masses & normal & $\langle \Mej\rangle=13 M_{\odot}$; $\sigma_\Mej=3\Msun$; (truncated at $20\Msun$) \\
\hline
\multicolumn{3}{l}{\textbf{Particle spectrum at injection}}\\
% \hline
Break energy & log-normal & $\langle E_b\rangle \simeq 0.28$ TeV;  $\,\,\sigma_{E_b} \simeq 0.12$ TeV \\
Low energy index & uniform & $1.0<\alpha_1<1.7$ \\
High energy index & uniform & $2.0<\alpha_1<2.7$ \\
Magnetic fraction & uniform & $0.02<\eta<0.2$ \\
%Containment factor & $\epsilon$ & $0.23$ \\
\hline
\end{tabular}
\end{table*}

\bibliographystyle{mn2e} 
\bibliography{pwn}

\end{document}